\documentclass{jpp}
\usepackage{graphicx}

\usepackage[utf8]{inputenc}
\usepackage[T1]{fontenc}
\usepackage{amsmath}

\shorttitle{Guidelines for authors}
\shortauthor{C. Granier, D. Borgogno, D. Grasso, E. Tassi}

\title{Gyrofluid analysis of electron $\bee$ effects on collisionless reconnection}

\author{C. Granier \aff{1,2}
  \corresp{\email{camille.granier@oca.eu}},
  D. Borgogno\aff{2}, D. Grasso\aff{2}
 \and E. Tassi \aff{1}}

\affiliation{\aff{1}Universit\'e C\^ote d'Azur, CNRS, Observatoire de la C\^ote d'Azur, Laboratoire J. L. Lagrange, Boulevard de l'Observatoire, CS 34229, 06304 Nice Cedex 4, France
\aff{2}Istituto dei Sistemi Complessi - CNR and Dipartimento di Energia, Politecnico di Torino, Torino 10129, Italy}

\begin{document}

\maketitle

\begin{abstract}
The linear and nonlinear evolutions of the tearing instability in a collisionless plasma with a strong guide field are analyzed on the basis of a two-field Hamiltonian gyrofluid model. The model is valid for a low ion temperature and a finite $\bee$. The finite $\bee$ effect implies a magnetic perturbation along the guide field direction and electron finite Larmor radius effects. A Hamiltonian derivation of the model is presented. A new dispersion relation of the tearing instability is derived for the case $\bee=0$ and tested against numerical simulations. For $\beta_e \ll 1$ the equilibrium electron temperature is seen to enhance the linear growth rate, whereas we observe a stabilizing role when electron finite Larmor radius effects become more relevant.  In the nonlinear phase, a double "faster-than-exponential" growth is observed, similarly to what occurs in the presence of ion finite Larmor radius effects.
Energy transfers are analyzed  and the conservation laws associated with the Casimir invariants of the model are also discussed.
Numerical simulations seem to indicate that finite $\beta_e$ effects do not produce qualtitative modifications in the structures of the  Lagrangian invariants associated with Casimirs of the model.
\end{abstract}

\section{Introduction}
Magnetic reconnection plays a crucial role in a broad range of plasma environments, from laboratory plasma experiments to astrophysical plasmas. It is a fundamental energy conversion process, as a result of which magnetic field energy is converted into kinetic energy and heat. In a reconnection even, the tearing instability is believed to play an important role as an onset mechanism of the process. A considerable progress in the understanding of this mechanism has been achieved through the fluid description of plasmas. The fluid framework is less costly in terms of computational resources, and physically more intuitive when compared to the kinetic framework. Fluid models, in general, are also more suitable for analytical treatment.  In the non-collisional case, some reduced fluid models were designed to retain two-fluid effects (e.g. \cite{Sch94, Gra99, Gra15, Del06, Fit07}), such as, for instance, electron inertia which is known to develop a thin current layer where modifications of the topology of the magnetic field lines can occur. These fluid models, on the other hand, neglect the effects of the electron Larmor radius, which makes it impossible to describe phenomena taking place at a microscopic scale comparable to that of the electron thermal gyro-radius. Gyrofluid models are the effective tools to fill this gap. Indeed, although obtained by truncating the infinite hierarchy of equations evolving the moments of the gyrokinetic equations, gyrofluid models, unlike fluid models, retain finite Larmor radius (FLR) effects and are thus valid on thermal Larmor radius scales. Also, most of the reduced fluid models neglect the perturbations of the magnetic field along the direction of a guide field, the latter typically corresponding to the mean magnetic field in astrophysical plasmas (e.g. \cite{Sch09}) or to an imposed external field in laboratory plasmas. However, even in the case of a strong guide field, such perturbations can be relevant in some nearly collisionless environments such as the solar wind, which motivates their inclusion in an analysis of collisionless reconnection.

In this work, we make use of a gyrofluid model to study the linear and nonlinear evolution of the tearing instability in a collisionless plasma with strong guide field. This study is based on a two-field gyrofluid model that has been derived from gyrokinetic equations in \cite{Tas20}, assuming a quasi-static closure. 
With respect to the above mentioned reduced fluid models, such gyrofluid model accounts for both finite electron Larmor radius effects and perturbations parallel to the direction of the guide field. The model is taken within the asymptotic cold ion limit, although we present a small set of simulations performed in the limit of hot ions to reflect the differences and possible consequences of this limit. A more in-depth study of the hot ion limit could be done in a subsequent work. Our gyrofluid model is valid for finite $\bee$ values, where $\bee$ is the ratio between the electron pressure and the magnetic pressure based on the guide field. 
We remark that finite $\beta_e$ effects were taken into account also in the model by \cite{Fit04, Fit07}. However, in that model, electron FLR effects were neglected. 
The study of reconnection for a finite $\bee$ can be relevant especially for astrophysical plasmas with large temperatures, such as in the Earth magnetosheath, where some $\beta >1$ values are observed, in the presence of a guide field, during reconnection events (\cite{Man20, Eas18}).

We consider magnetic reconnection taking place in a two dimensional $(2D)$ plane, perpendicular to the guide field component. Reconnection is mediated by electron inertia and by electron FLR, which makes the process non-dissipative, unlike reconnection driven by electrical collisional resistivity. As many dissipationless fluid and gyrofluid models, also the gyrofluid model under consideration possesses a Hamiltonian structure, which reveals the presence of two Lagrangian invariants and gives the expression of the conserved total energy of the system. With this we can obtain further information about how $\bee$ can influence the distribution of the different components of the total energy. 

In the limit $\bee \rightarrow 0$ (in the following also referred to as the "fluid" limit), the model corresponds to the two-field fluid model of \cite{Sch94}. This fluid model has long been used to study the tearing instability, and a relevant dispersion relation for the collisionless tearing mode, applicable to this model, has been derived in \cite{Por91}. We present in this article a new analytical formula, valid assuming the constant-psi approximation (\cite{Fur63}), which differs from the relation of \cite{Por91}, taken in the limit where the tearing stability parameter $\Delta '$ is small, by the presence of a small corrective term. These two formulas are tested against numerical simulations and, in its regime of validity, our new relation shows a better agreement with the numerical growth rate.

We studied numerically the effect of a finite $\bee$ in the linear and nonlinear phase of the tearing instability. For the linear phase, we first isolate the effect of varying $\beta_e$ by keeping fixed all the other parameters of the system. In this setting we observe a stabilizing role of the $\beta_e$ parameter. The stabilizing effect is then seen to be reduced when increasing  the normalized electron skin depth $d_e$. A partial justification of this behavior can be given analytically considering the small FLR limit of the model. We remark that varying $\beta_e$ with fixed $d_e$ and $\rho_s$ amounts to varying the normalized thermal electron Larmor radius $\rho_e$ at fixed $\rho_s$. 
Subsequently, we consider the effect of varying $\beta_e$ while keeping a fixed mass ratio.  The previously mentioned stabilizing role of $\beta_e$ is then concomitant with the destabilizing role of the normalized sonic Larmor radius $\rho_s$.  The growth rate is thus evaluated for different values of the parameters $d_e$, $\rs$ and $\rho_e$. These parameters are associated with different physical scales and are absent in the usual reduced magnetohydrodynamics (MHD) description.  The results we find turn out to be in agreement with those of \cite{Num11} and of \cite{Num15}, which were obtained with a gyrokinetic model. In the nonlinear phase, we find the explosive growth rate (\cite{Ayd92}) which has been obtained as well in the gyrofluid study of \cite{Bia12} that was considering low $\bee$ and ion FLR but no electron FLR effects. We investigate how the effects of $\bee$ affects this faster than exponential growth.

The reconnection process described by Hamiltonian reduced fluid and gyrofluid models has been analyzed in terms of Lagrangian invariants in several cases in the past (\cite{Caf98, Gra01, Gra10, Com13, Gra15}). The effect of both electron FLR effects and parallel magnetic perturbations on the structure of such invariants has not been studied so far, though.
In this paper we present the behavior of the two topological invariants of the system. They extend the Lagrangian invariants of simpler models that do not account for $\bee$ effects and behave similarly. 

 The paper is organized as follows. In Sec. \ref{sec:model} we derive the gyrofluid model adopted for the analysis. The procedure we follow for the derivation automatically provides the Hamiltonian structure of the model. Section \ref{sec:lin} contains a review of the linear theory and a new dispersion relation for the case $\bee=0$. We also present the results of numerical simulations in the linear phase, for arbitrary $\bee$. In Sec. \ref{sec:nonlin} the results obtained in the non-linear phase are presented and the gyrofluid version is compared to the fluid version. In this Section, we also study the impact of a finite $\bee$ on the evolution of the different energy components.  Section \ref{sec:inv} presents the conservation laws and the evolution of the Lagrangian invariants of the model. In the Appendix we present the derivation of the new dispersion relation, which is based on the asymptotic matching theory.

\section{The gyrofluid model}\label{sec:model}

We begin by considering the model given by the evolution equations

\begin{align}
& \frac{\pa N_i}{\pa t}+[\gamui \phi + \taupi \rspe^2 2 \gamdi \bpar , N_i]-[\gamui \apar , U_i]=0,  \label{conti}\\
& \frac{\pa}{\pa t}(U_i + \gamui \apar) + [\gamui \phi + \taupi \rspe^2 2 \gamdi \bpar , U_i + \gamui \apar]-\frac{\taupi \rspe^2}{\Theta_i} [ \gamui \apar , N_i]=0,  \label{momi}\\
&\frac{\pa N_e}{\pa t}+[\gamue \phi - \rspe^2 2 \gamde \bpar , N_e]- [\gamue \apar , U_e]=0,  \label{conte}\\
&\frac{\pa}{\pa t}(\gamue \apar - d_e^2 U_e)+[\gamue \phi - \rspe^2 2 \gamde \bpar , \gamue \apar - d_e^2 U_e]+\frac{\rspe^2}{\Theta_e}[\gamue \apar ,N_e]=0,  \label{mome}
\end{align}
complemented by the static relations
\begin{align}
& \gamui N_i - \gamue N_e + (1-\Theta_i)\gammzi \frac{\phi}{\taupi \rspe^2} +  (1-\Theta_e)\gammze \frac{\phi}{ \rspe^2} + (\Theta_i\gamui ^2 -1)\frac{\phi}{\taupi \rspe^2}\nno \\ 
& + (\Theta_e\gamue^2 -1) \frac{\phi}{\rspe^2} + (\Theta_i\gamui 2 \gamdi -\Theta_e\gamue 2 \gamde)\bpar \nno \\ 
& + ((1- \Theta_i) (\gammzi - \gammui) - (1 - \Theta_e) (\gammze - \gammue))\bpar =0,  \label{qn4f}\\
& \lapp \apar = \left( \left(1 - \frac{1}{\Theta_e}\right)(1 - \gammze)\frac{1}{d_e^2}  + \left(1 - \frac{1}{\Theta_i}\right)(1 - \gammzi)\frac{1}{d_i^2}   \right)\apar \nno \\ 
& + \gamue U_e - \gamui U_i,   \label{amppar4f}\\
& \bpar =-\frac{\bepe}{2}\left(\taupi 2 \gamdi N_i +   2 \gamde N_e    +   (1-\Theta_i)( \gammzi- \gammui) \frac{\phi}{ \rspe^2} \right. \nno \\ 
& - (1-\Theta_e)( \gammze- \gammue) \frac{\phi}{ \rspe^2}  +   \Theta_i \gamui 2 \gamdi \frac{\phi}{\rspe^2}  - \Theta_e \gamue 2 \gamde \frac{\phi}{\rspe^2}   + \Theta_i \taupi 4 \gamdi^2 \bpar   \nno \\
& \left. + \Theta_e 4 \gamde^2 \bpar   + \taupi 2(1- \Theta_i)( \gammzi- \gammui)\bpar  + 2(1- \Theta_e)( \gammze- \gammue)\bpar \right)  \label{ampperp4f}
\end{align}
Equations (\ref{conti}) and (\ref{conte}) correspond to the ion and electron gyrocenter continuity equations, respectively, whereas Eqs. (\ref{momi}) and (\ref{mome}) refer to the ion and electron momentum conservation laws, along the guide field direction.

The static relations (\ref{qn4f}), (\ref{amppar4f}) and (\ref{ampperp4f}) descend from quasi-neutrality and from the projections of Amp\`ere's law along directions parallel and perpendicular to the guide field, respectively.

The system (\ref{conti})-(\ref{ampperp4f}), although written with a different normalization, consists to the Hamiltonian
four-field model derived by \cite{Tas20}, taken in the 2D limit (assuming that all the independent variables do not vary along the direction of the guide field). This model has been derived by considering a quasi-static closure which fixes all the moments, except for the gyrocenter density and velocity parallel to the guide field, for both species. Strictly speaking, the derivation of the quasi-static closure, followed by \cite{Tas20}, does not hold in the purely 2D case, which we consider then as a limit of the 3D case as the component of the wave-vector of the perturbation along the guide field, goes to zero. We recall that the quasi-static closure adopted by \cite{Tas20} is valid in 3D, when, for each particle species, the phase velocity of the fluctuations along the guide field direction is much less than the thermal speed based on the parallel equilibrium temperature of the corresponding species.

The model is formulated in a slab geometry adopting a Cartesian coordinate system $(x,y,z)$. We indicated with $N_s$ and $U_s$ the fluctuations of the gyrocenter densities and velocities parallel to the guide field, respectively, for the species $s$, with $s=e$ for electrons and $s=i$ for ions. The symbols $\apar, \bpar$ and $\phi$, on the other hand, corresponds to the fluctuations of the $z$ component of the magnetic vector potential, to the parallel magnetic perturbations and to the fluctuations of the electrostatic potential, respectively. 
The fields $N_{e,i}, U_{e,i}, \apar, \bpar$ and $\phi$ depend on the time variable $t$ and on the spatial coordinates $x$ and $y$, which belong to the domain $\mathcal{D}=\{-L_x \leq x \leq L_x \, , \, -L_y \leq y \leq L_y \}$, with $L_x$ and $L_y$ positive constants. Periodic boundary conditions are imposed on the domain $\mathcal{D}$. The operator $[ \, , \, ]$ is the canonical Poisson bracket and is defined by $[f,g]=\partial_x f \partial_y g - \partial_y f \partial_x g$, for two functions $f$ and $g$.

We write the normalized magnetic field in the form
\begin{equation} \label{magfield}
    \mathbf{B}(x,y,z,t) \approx  \hat{z}+ \frac{\hat{d}_i}{L}\bpar(x,y,z,t)\mathbf{z} + \nabla \apar (x,y,z,t)\times \hat{z},
\end{equation}
with $\hat{z}$ indicating the unit vector along the $z$ direction, with $L$ a characteristic equilibrium scale length, and with $\hat{d}_i=c\sqrt{m_i/(4 \pi e^2 n_0)}$ the ion skin depth. We denote by $m_i$ the ion mass, by $e$ the proton charge, by $c$ the speed of light and $n_0$ the equilibrium density (equal for ions and electrons).  
The first term on the right-hand side of (\ref{magfield}) accounts for the strong guide field. In Eq. (\ref{magfield}) only up to the first order terms in the fluctuations are shown, and the higher-order contributions, which guarantee $\nabla \cdot \mathbf{B}=0$, are neglected.
The normalization of the variables used in Eqs. (\ref{conti})-(\ref{ampperp4f}) is the following:
\begin{align}
& t=\frac{v_A}{L}\hat{t}, \qquad x=\frac{\hat{x}}{L}, \qquad y=\frac{\hat{y}}{L}, \\
& N_{e,i}=\frac{L}{\hat{d}_i}\frac{\hat{N}_{e,i}}{n_0}, \qquad U_{e,i}=\frac{L}{\hat{d}_i}\frac{\hat{U}_{e,i}}{v_A},\\
& \apar=\frac{\hat{A}_\parallel}{L B_0}, \qquad \bpar=\frac{L}{\hat{d}_i}\frac{\hat{B}_\parallel}{B_0}, \qquad \phi=\frac{c}{v_A} \frac{\hat{\phi}}{L B_0},
\end{align}
where the hat indicates dimensional quantities, $B_0$ is the amplitude of the guide field and $v_A=B_0/\sqrt{4 \pi m_i n_0}$ is the Alfv\'en speed.  

Independent parameters in the model are $\bepe$, $\taupi$, $\rspe$, $\Theta_e$, $\Theta_i$ and $d_e$, corresponding to the ratio between equilibrium electron pressure and magnetic guide field pressure, to the ratio between equilibrium perpendicular ion and electron temperatures, to the normalized sonic Larmor radius, to the ratio between the equilibrium perpendicular and parallel temperature for electrons and ions and to the normalized perpendicular electron skin depth, respectively. These parameters are defined as
\begin{align}
& \bepe=8 \pi \frac{n_0 \Tpee}{B_0^2}, \qquad \taupi=\frac{\Tpei}{\Tpee}, \qquad \rspe=\frac{1}{L}\sqrt{\frac{\Tpee}{m_i}}\frac{m_i c}{e B_0}, \\
& \Theta_e= \frac{\Tpee}{\Tpae},\qquad \Theta_i= \frac{\Tpei}{\Tpai}, \qquad d_e=\frac{1}{L}c \sqrt{\frac{m_e}{4 \pi e^2 n_0}},
\end{align}
where $\Tpea$ and $\Tpa$ are the perpendicular and parallel equilibrium temperatures for the species $s$, respectively, and $m_e$ is the electron mass. Note that $\rspe/\sqrt{\bepe / 2}=d_i$, where $d_i=\hat{d}_i /L$ is the normalized ion skin depth.

Electron and ion gyroaverage operators are associated with corresponding Fourier multipliers in the following way:
\begin{align}
&\gamue=2\gamde \rightarrow \mathrm{e}^{-\kpe \frac{\bepe}{4}d_e^2}, \label{op1}\\
&\gamui=2\gamdi \rightarrow \mathrm{e}^{-\kpe \frac{\taupi}{2}\rspe^2}. \label{op2}
\end{align}
and 
\beq \label{gammaoperatore}
\Gamma_{0e}  \rightarrow  I_0\left(\kpe \frac{\bepe}{2}d_e^2\right) e^{-\kpe \frac{\bepe}{2}d_e^2} , \qquad \Gamma_{1e}   \rightarrow I_1\left(\kpe \frac{\bepe}{2}d_e^2\right) e^{-\kpe \frac{\bepe}{2}d_e^2},
\eeq
\beq \label{gammaoperatori}
\Gamma_{0i}  \rightarrow  I_0\left(\kpe \taupi \rspe^2 \right) e^{-\kpe \taupi \rspe^2} , \qquad \Gamma_{1i}   \rightarrow I_1\left(\kpe \taupi \rspe^2 \right) e^{-\kpe \taupi \rspe^2},
\eeq
where $I_n$ are the modified Bessel functions of order $n$ and $ \kpe = \sqrt{k_{x}^{2}+ k_{y}}^{2} $ is the perpendicular wave number.\\
For the range of parameters adopted in our analysis, the gyroaverage operators $\gamue$ and $\gamui$, corresponding to those introduced by \cite{Bri92},  are shown to be adequate.
Nevertheless, different gyroaverage operators, described in the papers \cite{Dor93}, \cite{Man18}, have proven to provide a very good agreement with the linear kinetic theory for a wider range of scales and are widespread in gyrofluid numerical codes. \\
We define the dynamical variables
\beq   \label{cmom}
A_i=\gamui \apar + d_i^2 U_i, \qquad A_e=\gamue \apar - d_e^2 U_e.
\eeq
The fields $A_i$ and $A_e$ are proportional to the parallel canonical fluid momenta, based on gyroaveraged magnetic potentials.

The two static relations (\ref{qn4f}) and (\ref{ampperp4f}) can be seen, in Fourier space, as an inhomogeneous linear system with the Fourier coefficients of $\phi$ and $\bpar$ as unknowns, for given $N_{i,e}$. From the solution of this system, one can express the fields $\phi$ and $\bpar$ in terms of $N_i$ and $N_e$, by means of relations of the form
\beq
\bpar=\call_B (N_i, N_e), \qquad \phi=\call_\phi (N_i , N_e),
\eeq
where $\call_B$ and $\call_\phi$ are linear operators, the explicit form of which can easily be provided in Fourier space. Similarly, using  the relations (\ref{amppar4f}) and
 (\ref{cmom}), one can express $U_e$ and $U_i$ in the form
 \beq
 U_e=\lue (A_i , A_e), \qquad U_i=\lui (A_i , A_e),
 \eeq
 where $\lue$ and $\lui$ are also linear operators.
 
 The model (\ref{conti})-(\ref{ampperp4f}) can be formulated as an infinite dimensional Hamiltonian system, adopting as dynamical variables the four fields $N_i$, $N_e$, $A_i$ and $A_e$ \citep{Tas20}.
 
 The corresponding Hamiltonian structure consists of the Hamiltonian functional
 \begin{align}
 &H(N_i , N_e , A_i ,A_e )=\frac{1}{2} \int d^2 x \, \left( \frac{\taupi \rspe^2}{\Theta_i} N_i^2 + \frac{ \rspe^2}{\Theta_e}  N_e^2 + A_i \lui (A_i , A_e)\right. \nno\\
 & \left. - A_e \lue (A_i , A_e)  + N_i (\gamui \call_\phi (N_i , N_e) + \taupi \rspe^2 2 \gamdi \call_B (N_i , N_e) )  \right. \nno\\ 
 & \left. - N_e ( \gamue \call_\phi (N_i , N_e) - \rspe^2 2 \gamde \call_B (N_i , N_e))\right),  \label{ham4f}
 \end{align}
 and of the Poisson bracket
 \begin{align}
&\{ F , G\}= -\int d^2 x \left( N_i \left([F_{N_i} ,  G_{N_i}]+\taupi \frac{2}{\bepe}\frac{\rspe^4}{\Theta_i} [F_{A_i}, G_{A_i}]\right) \right. \nno\\
&\left.+  A_i \left([F_{A_i} , G_{N_i}]+[F_{N_i} , G_{A_i}]\right) -N_e([F_{N_e} , G_{N_e}] + d_e^2 \frac{\rspe^2}{\Theta_e} [F_{A_e} , G_{A_e}]) \right. \nno\\
& \left.-A_e([F_{A_e} , G_{N_e}] + [F_{N_e} , G_{A_e}])\right),  \label{pb4f}
\end{align}
where subscripts on functionals indicate functional derivatives, so that, for instance, $F_{N_i}=\delta F / \delta N_i$. Using the Hamiltonian (\ref{ham4f}) and the Poisson bracket (\ref{pb4f}), the four equations (\ref{conti})-(\ref{mome}) can be obtained from the Hamiltonian form \citep{Mor98}
\beq
\frac{\pa \chi}{\pa t}=\{\chi , H \},
\eeq
 replacing $\chi$ with $N_i$, $N_e$, $A_i$ and $A_e$. This Hamiltonian four-field gyrofluid model, although greatly simplified with respect to the original gyrokinetic system, is still amenable to a further reduction, concerning in particular the ion dynamics which, for the analysis of reconnection of interest here, was shown not to be crucially relevant (\cite{Com13}, \cite{Num11}). Also, we carry out most of the analysis in the isotropic cold-ion limit, a simplifying assumption which is also helpful for the comparison with previous works. Nevertheless, some comments will be provided also with regard to the opposite limit of equilibrium ion temperature much larger than the electron one. On the other hand, in carrying out the reduction procedure, we find it important to preserve a Hamiltonian structure, which avoids the introduction of uncontrolled dissipation in the system and also allows for a more direct comparison with previous Hamiltonian models for reconnection, in particular with the two-field model considered by \cite{Caf98}, \cite{Gra01}, \cite{Del06}, \cite{Del03}.
In particular, we intend to obtain a Hamiltonian reduced version of the four-field model (\ref{conti})-(\ref{ampperp4f}), in which the gyrocenter ion density fluctuations $N_i$ and ion gyrocenter parallel velocity fluctuations $U_i$ are neglected, the ion equilibrium temperature is isotropic, and ions are taken to be cold. The latter four conditions amount to impose
\beq   \label{cond1}
N_i=0, \qquad U_i=0, \qquad \Theta_i=1, 
\eeq
and take the limit
\beq   \label{cond2}
\taupi \rightarrow 0.
\eeq
Because we want to perform this reduction while preserving a Hamiltonian structure, we apply the conditions (\ref{cond1}) and (\ref{cond2}) at the level of the Hamiltonian structure, instead of applying them directly to the equations of motion. The latter procedure would indeed produce no information about the Hamiltonian structure of the resulting model.

As first step, we impose the conditions (\ref{cond1})-(\ref{cond2}) in the static relations (\ref{qn4f})-(\ref{ampperp4f}), which leads to
 \begin{align}
&\left( \frac{(1-\Theta_e)}{\rspe^2}\gammze+ \frac{(\Theta_e\gamue^2-1)}{\rspe^2} + \lapp\right) \phi \nno  \\ 
& -\left(\Theta_e \gamue 2 \gamde-1+(1 -\Theta_e) (\gammze - \gammue) \right)\bpar =\gamue N_e,  \label{qncond}\\
&\left(\left(1-\frac{1}{\Theta_e}\right)\frac{(\gammze-1)}{d_e^2} + \lapp\right)\apar=  \gamue U_e,  \label{ampparcond}\\
&\left(\Theta_e \gamue 2 \gamde+ (1-\Theta_e)(\gammze-\gammue)-1\right)\frac{\phi}{\rspe^2} \nno \\ &-\left(\frac{2}{\bepe}+ 2(1-\Theta_e)(\gammze-\gammue)+ 4\Theta_e \gamde^2\right)\bpar=2 \gamde N_e.  \label{ampperpcond}
\end{align}
The three relations (\ref{qncond})-(\ref{ampperpcond}), together with the definition of $A_e$ in Eq. (\ref{cmom}), make it possible to express $\bpar$, $\phi$ and $U_e$, in terms of the two dynamical variables $N_e$ and $A_e$, according to
\beq
\bpar=\calbo N_e , \qquad \phi=\calphio N_e , \qquad U_e=\calueo A_e,
\eeq
where $\calbo$, $\calphio$ and $\calueo$ are linear symmetric operators.

As next step, we impose the conditions (\ref{cond1})-(\ref{cond2}) on the Hamiltonian (\ref{ham4f}), which reduces the Hamiltonian to the following functional of the only two dynamical variables $N_e$ and $A_e$:
 \begin{align}
 &H(N_e ,A_e )=\frac{1}{2} \int d^2 x \, \left( \frac{\rspe^2}{\Theta_e} N_e^2  - A_e \calueo  A_e  - N_e ( \gamue \calphio N_e - \rspe^2 2 \gamde \calbo  N_e)\right).  \label{ham2f}
 \end{align}
With regard to the Poisson bracket (\ref{pb4f}), we can consider  its limit as $\taupi \rightarrow 0$, given that the bilinear form (\ref{pb4f}) is a valid Poisson bracket for any value of $\taupi$. On the other hand, in general, we cannot impose directly the conditions (\ref{cond1}) in the bracket, as this operation does not guarantee that the resulting bilinear form satisfies the Jacobi identity. However, we remark that the set of functionals of the two dynamical variables $N_e$ and $A_e$, which the reduced Hamiltonian (\ref{ham2f}) belongs to, forms a sub-algebra of the algebra of functionals of $N_i$, $N_e$, $A_i$ and $A_e$, with respect to the Poisson bracket (\ref{pb4f}). Indeed, if $F$ and $G$ are two functionals of $N_e$ and $A_e$ only, $\{F , G \}$ is again a functional of $N_e$ and $A_e$ only. One can in particular restrict to the part of the bracket (\ref{pb4f}) involving functional derivatives only with respect to $N_e$ and $A_e$, the other terms yielding vanishing contributions when evaluated on functionals of $N_e$ and $A_e$ only. The resulting Poisson bracket therefore reads
 \begin{align}
&\{ F , G\}= \int d^2 x \left( N_e([F_{N_e} , G_{N_e}] +  d_e^2\frac{\rspe^2}{\Theta_e} [F_{A_e} , G_{A_e}]) +A_e([F_{A_e} , G_{N_e}] + [F_{N_e} , G_{A_e}])\right).  \label{pb2f}
\end{align}
We remark that the Poisson bracket (\ref{pb2f}) has the same form as that of the model investigated by \cite{Caf98} and by \citet{Gra01}.

The resulting reduced two-field model, accounting for the conditions (\ref{cond1})-(\ref{cond2}), can then be obtained from the Hamiltonian (\ref{ham2f}) and the Poisson bracket (\ref{pb2f}). The corresponding evolution equations read
\begin{align}
&\frac{\pa N_e}{\pa t}+[\gamue \phi - \rspe^2 2 \gamde \bpar , N_e]- [\gamue \apar , U_e]=0,  \label{conte2}\\
&\frac{\pa A_e}{\pa t}+[\gamue \phi - \rspe^2 2 \gamde \bpar , A_e]+\frac{\rspe^2}{\Theta_e}[\gamue \apar ,N_e]=0,  \label{mome2}
\end{align}
where $\bpar$, $\phi$ and $U_e$ are related to $N_e$ and $A_e$ by means of Eqs. (\ref{cmom}) and (\ref{qncond})-(\ref{ampperpcond}). 

We impose now electron temperature isotropy (i.e. setting $T_{0 \perp  e} =T_{0 \parallel e}=T_{0e}$, corresponding to $\Theta_e=1$)  and the evolution equations are reduced to 

\begin{align}
&\frac{\pa N_e}{\pa t}+[\gamue \phi - \rs^2 2 \gamde \bpar , N_e]- [\gamue \apar , U_e]=0,  \label{conteiso}\\
&\frac{\pa A_e}{\pa t}+[\gamue \phi - \rs^2 2 \gamde \bpar , A_e]+\rs^2[\gamue \apar ,N_e]=0,  \label{momeiso}
\end{align}
complemented by the equations 
 \begin{align}
&\left( \frac{ \gamue^2 -1}{\rs^2}+\lapp\right) \phi-\left(  \gamue 2 \gamde -1\right)\bpar =\gamue N_e,  \label{qncondiso}\\
&\lapp\apar=\gamue U_e,  \label{ampparcondiso}\\
&\left( \gamue 2 \gamde-1\right)\frac{\phi}{\rs^2} -\left(\frac{2}{\bee}+ 4 \gamde^2\right)\bpar=2 \gamde N_e.  \label{ampperpcondiso}
\end{align}
Eqs. (\ref{conteiso}), (\ref{momeiso}) and (\ref{qncondiso})-(\ref{ampperpcondiso}) correspond to the gyrofluid model adopted for the subsequent analysis of magnetic reconnection.


\section{Linear phase}\label{sec:lin}

\subsection{Linear theory for $\bee \rightarrow 0$}

In this Subsection we focus on the regime for which the electron FLR effects and the parallel magnetic perturbations are negligible. The limit of vanishing thermal electron Larmor radius, i.e. $\rho_e=d_e \sqrt{\bee /2} \rightarrow 0$, is adopted by considering $\bee \rightarrow 0$ and a fixed $d_e$. 
This limit enables to reduce the gyrofluid model  (\ref{conteiso})-(\ref{ampperpcondiso}) to the fluid model of \cite{Sch94, Caf98}, for which the analytical study of the tearing instability has been extensively studied in the past (\cite{Por91, Gra01, Gra99}).\\
When assuming $\bee \rightarrow 0$ for a fixed $d_e$, the gyroaverge operators can be approximated in the Fourier space in the following way
\beq
\begin{split} \label{approxG2}
& \gamue f(x,y) = \left(1 +\rho_e^2 \lapp\right) f(x,y) + O(\rho_e^4), \\
&\gamde f(x,y) = \frac{1}{2}\left( 1 + \rho_e^2 \lapp \right) f(x,y) + O(\rho_e^4).
\end{split}
\eeq 
Using this development in Eqs. (\ref{conteiso})-(\ref{ampperpcondiso}) and neglecting the first order correction, we obtain the evolution equations (\cite{Sch94})
\begin{equation} \label{fluid1}
  \frac{\partial \lapp \phi}{\partial t} + [\phi, \lapp \phi] - [ \apar, \lapp \apar] = 0,
\end{equation} 
\begin{equation} \label{fluid2}
\begin{split} 
\frac{\partial}{\partial t} \left( \apar - d_e^2 \lapp \apar\right) + \left[\phi , \apar - d_e^2  \lapp \apar\right] - \rho_s^2[\lapp \phi, \apar] =0.
\end{split}
\end{equation}
We assume an equilibrium given by 
\beq \label{equilibrium}
\phi^{(0)}(x) = 0, \qquad \apar^{(0)} (x)=  \frac{\lambda}{\cosh^2 \left( \frac{x}{\lambda} \right)},
\eeq
where $\lambda$ is a parameter that stretches the equilibrium scale length and modifies the equilibrium amplitude.  %
We consider the perturbations
\beq \label{perturbations}
\apar^{(1)} (x,y,t) = \wapar (x) e^{\gamma t +i k_y y} + \bar{\tilde{A}} (x) e^{\gamma t -i k_y y} , \quad \phi^{(1)}(x,y,t) = \wphi(x) e^{\gamma t +i k_y y} + \bar{\tilde{\phi}}(x) e^{\gamma t -i k_y y} ,
\eeq
where $\gamma$ is the growth rate of the instability, $k_y = \pi m/ L_y$ is the wave number, with $m \in \mathbb{N}$ and the overbar refers to the complex conjugate.
The collisionless tearing mode has been studied in \cite{Por91} for the $m=1$ mode in toroidal geometry and the results can be adapted to the model (\ref{fluid1})-(\ref{fluid2}). In particular, a dispersion relation has been obtained analytically and is valid for small and large values of the tearing stability parameter $\Delta'$, with
\begin{equation} \label{Delta'}
    \Delta' =  \lim_{x \rightarrow 0^{+}} \frac{\wapar_{out}'}{\wapar_{out}}  - \lim_{x \rightarrow 0^{-}} \frac{ \wapar_{out}'}{\wapar_{out}},
\end{equation}
where $\tilde{A}_{out}$ is the solution for $\tilde{A}$ of the linearized system in the outer region (see also the Appendix).  The tearing index, $\Delta'$, is a common measure of the discontinuity of the logarithmic derivative of $\wapar_{out}$ at the resonant surface. 
The dispersion relation is given by (\cite{Por91}, \cite{Fit10})
\begin{equation}
    \frac{\pi}{2}\left(\frac{\gamma }{2 k_y}\right)^2 = - \rho_s \frac{\pi}{\Delta'} + \rho_s^2 d_e  \frac{2 k_y}{\gamma}. \label{dispPorcelli}
\end{equation}

In the limit $d_e^{2/3} \rho_s^{1/3} \Delta'\ll 1$, the relation (\ref{dispPorcelli}) is reduced to 
\beq \label{smalldelta'}
\gamma = 2 k_y \frac{d_e \rho_s}{\pi } \Delta'.
\eeq
In the Appendix of this paper, we present the derivation of a new dispersion relation valid in the limit $ (\gamma d_e/ (k_y \rho_s))  \Delta' \ll 1$. In the appropriate regime of validity, the new dispersion relation includes a corrective term to Eq. (\ref{smalldelta'}). We derived this dispersion relation using an asymptotic matching method and various assumptions, slightly different from those adopted by \cite{Por91}. 
\begin{table}
\caption{Table summarizing the various assumptions} 
\begin{tabular}{llll}
 \hline\hline \\ \vspace{0.1cm}
 No. &  Assumptions used \\  \hline \\    \vspace{0.2cm}
 1& Time variation of the perturbation is slow   &  $\frac{\gamma}{k_y}\ll 1$ \\  \vspace{0.2cm} 
 2& Smallness of the inner scales   &  $\frac{\gamma d_e}{k_y \rho_s} \ll \rho_s \ll 1$   \\    \vspace{0.2cm}
 3& Use of the constant $\psi$ approximation   &   $\frac{\gamma d_e}{k_y \rho_s}\Delta' \ll 1$      \\\vspace{0.2cm}
 4& Neglecting FLR effects in the inner regions & $ \rho_e \ll \frac{\gamma d_e}{k_y \rho_s}$,    \\   
 \hline\hline \vspace{0.1cm}
\end{tabular}
\end{table}
Table 1 gives a review of the assumptions that were adopted on the parameters during our the analysis.  The assumption No. 1 indicates a slow time variation of the perturbation. The No. 2 is the assumption on the scales of the inner region, where electron inertia becomes important and allows the break of the frozen flux condition. The assumption No. 3 allows the use of the so-called \textit{constant $\psi$ approximation}, implying that the dispersion relation is valid for large wave numbers (\cite{Fur63}). The condition 4, imposed to neglect electron FLR, can be verified for a low-$\bee$ plasma. From a technical point of view, our new dispersion relation is obtained by solving the equations in the inner layer in real space, unlike in \cite{Por91} where the corresponding equations are transformed and solved in Fourier space.
The result of our linear theory, which is described in more detail in the Appendix, is given by the dispersion relation,
 \beq \label{dispcorrected}
\gamma = 2 k_y \frac{d_e \rho_s}{\pi \lambda} \Delta' + \frac{\gamma^2 d_e \pi \lambda}{4 k_y \rho_s^2}.
\eeq
The first term in the right hand side of (\ref{dispcorrected}) is exactly that of the formula (\ref{smalldelta'}), for $\lambda=1$. In the parameter regime indicated by Table 1, the second term in (\ref{dispcorrected}) is a small term that provides a correction to the formula (\ref{smalldelta'}).\\
A solution of the dispersion relation (\ref{dispcorrected}), considered in the regime identified by the assumptions of Table 1, is   
\beq 
\gamma_u = 2 k_y \left( \frac{ \rho_s ^2}{\pi  d_e  \lambda }-\frac{\rho_s^{3/2}\sqrt{\rho_s - 2 d_e^2 \Delta'}}{\pi 
   d_e  \lambda } \right), \label{corrected}
\eeq
and is real for $ \rho_s > 2 d_e^2 \Delta'$. 

This new dispersion relation is tested against numerical simulations and compared to the expression (\ref{smalldelta'}).
The numerical solver is pseudo-spectral and is based on a third order Adam-Bashforth scheme. The scheme uses numerical filters acting on typical length scales much smaller than the physical scales of the system (\cite{Lel92}). The instability is triggered by perturbing the equilibrium with a disturbance of the parallel electron gyrocenter velocity field.  Because of the requirement of periodic boundary conditions, the equilibrium (\ref{equilibrium}) is approximated by
\beq
\apar^{(0)} (x)=\sum^{30}_{n=-30} a_n e^{i n x},
\eeq
where $a_n$ are the Fourier coefficients of the function $f(x)= \lambda/\cosh \left(\frac{x}{\lambda} \right)^2$ (\cite{Gra06}).
The numerical growth rate is determined by the formula 
\beq \label{numericalgrowthrate}
\gamma_N = \frac{d}{dt} \log\left| \apar^{(1)} \left(\frac{\pi}{2},0,t \right)  \right|,
\eeq
so that $A_\parallel^{(1)}$ is evaluated at the $X$-point, where reconnection takes place.\\

\begin{figure}
    \centering
\includegraphics[trim=0 0 0 0, scale=0.7]{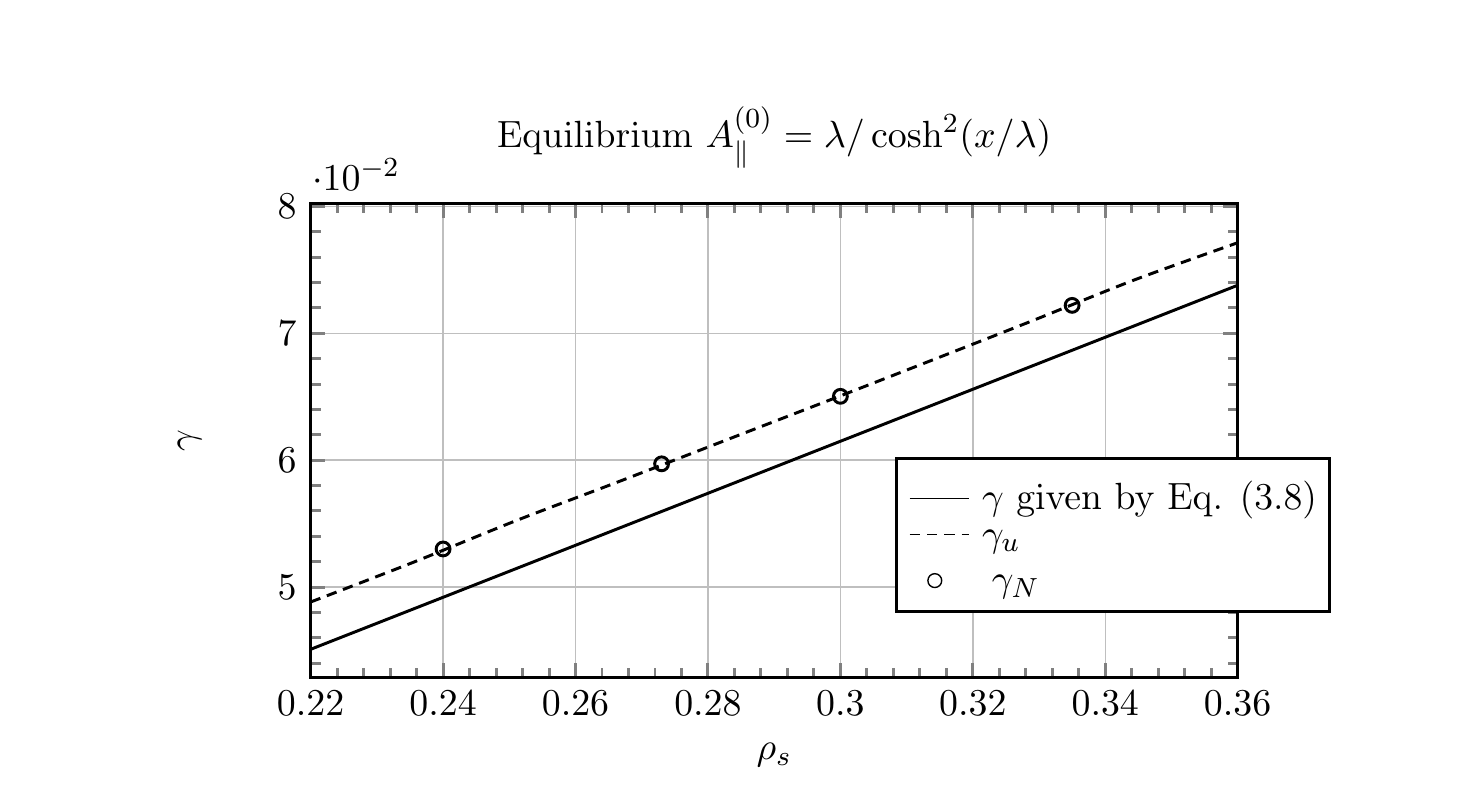}
    \caption{Comparison between the analytical growth rate $\gamma_u$ obtained from the new formula (\ref{corrected}) (dashed line), the analytical growth rate obtained from the formula (\ref{smalldelta'}) (solid line) and the numerical growth rate $\gamma_N$ defined in Eq. (\ref{numericalgrowthrate}) (circles). The parameters are $d_e=0.1$, $\lambda=1$, $\Delta'=0.72$, $m=1$. The box size is given by $- 10\pi < x < 10\pi$, $- 0.48\pi< y < 0.48\pi$. The values of the parameters lie in the regime of validity of the new formula (\ref{corrected}). One can see that, for different values of $\rho_s$, the correction present in Eq. (\ref{corrected}) yields a better agreement with the numerical values.}
    \label{fig:gtc}
\end{figure}
\begin{figure}
    \centering
 
\includegraphics[trim=0 0 0 0, scale=0.68]{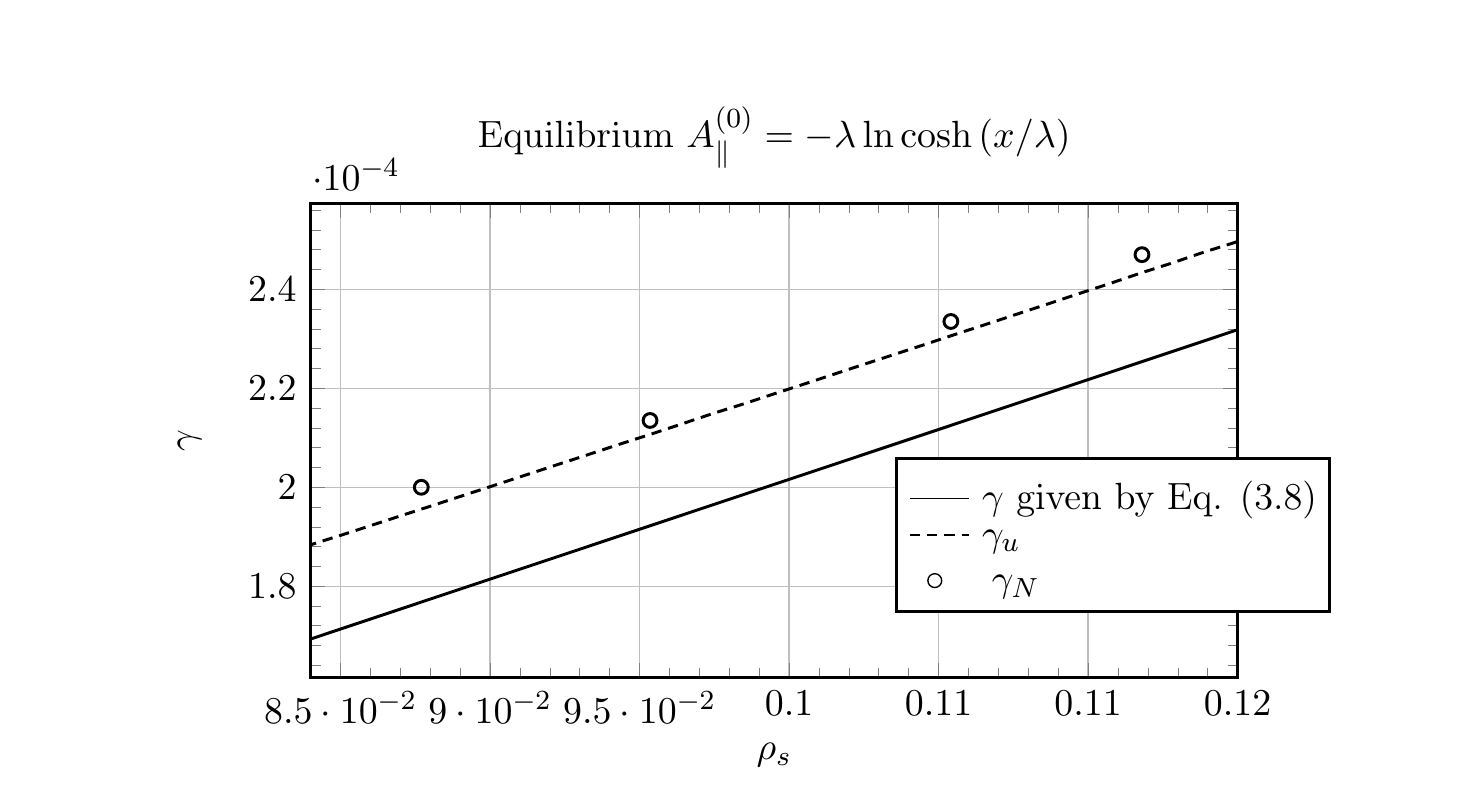}
    \caption{This plot is showing additional tests, analogous to those of Fig. 1, but with the Harris sheet equilibrium $\apar^{(0)} (x)= - \lambda \ln \cosh (x/\lambda)$, and $\phi^{(0)}(x) = 0$, for which $\Delta' = \frac{2}{\lambda}\left( \frac{1}{k_y \lambda} - k_y \lambda\right)$ and using the mode $m=1$. The parameters are $d_e=0.2$ and $\lambda=3$. The box size is $- 10\pi < x < 10\pi$, $- 4\pi< y < 4\pi$. For this case, $\Delta'=0.38$. For this equilibrium the dispersion relation determining $\gamma_u$ corresponds to Eq. (\ref{corrected}) with the right-hand side multiplied by a factor $1/2$. Symbols are the same as in Fig. 1. Also in this case, the new formula (\ref{corrected}) yields a better agreement with the numerical values. }
    \label{fig:gth}
\end{figure}
As shown in Figs. \ref{fig:gtc} and \ref{fig:gth}, the agreement between the theoretical and the numerical values appears to be improved by this new formula, when the latter is applied in its regime of validity. 
We also performed additional tests on a different equilibrium (the Harris sheet), as shown on Fig. \ref{fig:gth}. Also in this case, we observe that our new dispersion relation provides a better agreement with the numerical values. 

\begin{figure}
    \centering
\includegraphics[trim=0 0 0 0, scale=0.8]{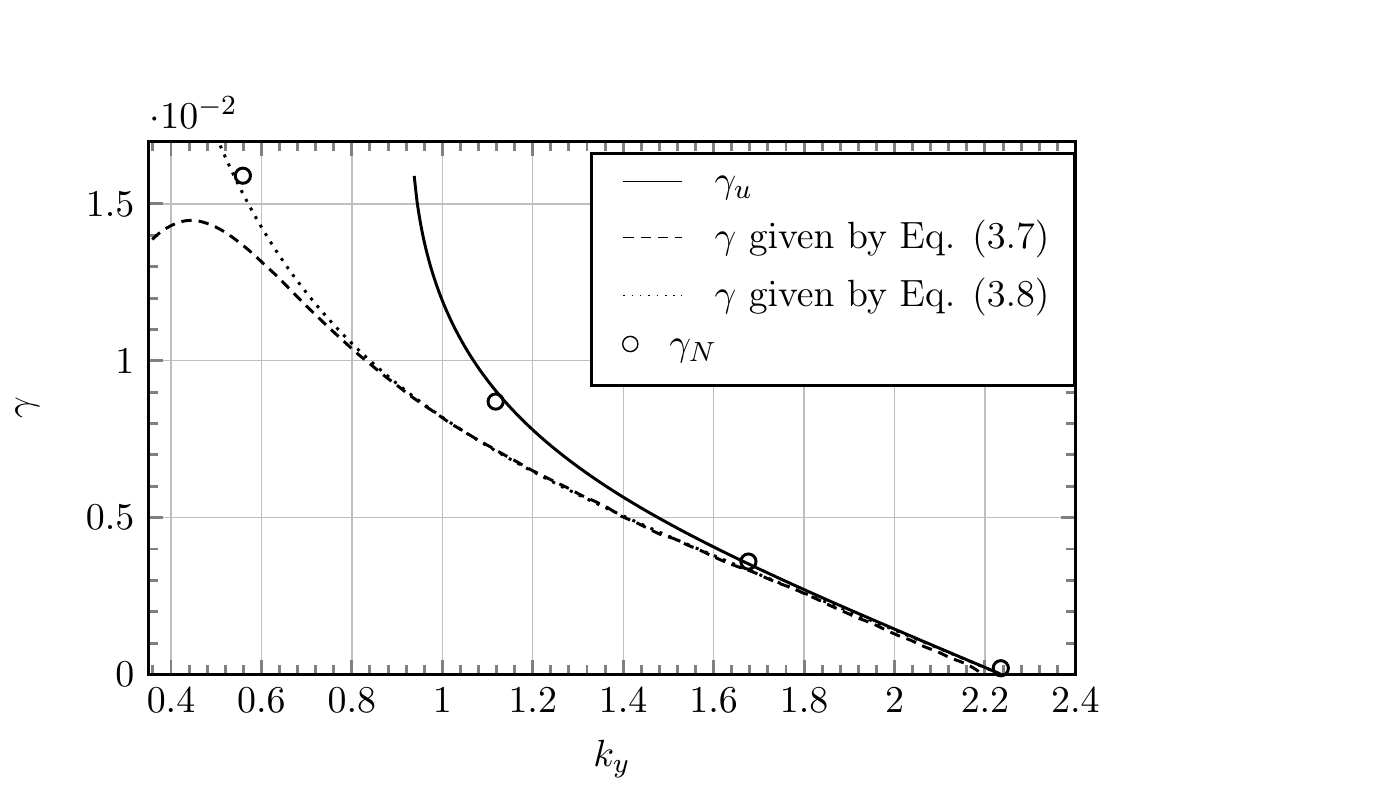}
    \caption{ Comparison between the theoretical growth rate predicted by Eqs. (\ref{dispPorcelli}), (\ref{smalldelta'}) and (\ref{corrected}), and the numerical growth rate $\gamma_N$. The parameters are $d_e=0.03$, $\rho_s=0.03$, $\lambda=1$.  The runs were done with the modes $1 \leq m \leq 4$. The box along $y$ is $1.789 \pi < y < 1.789 \pi$. The corresponding values of the tearing stability parameter lie in the interval  $0.005 \leq \Delta' \leq 47.86$. }
    \label{fig:ky}
\end{figure}

Consequently, (\ref{corrected}) can be seen as an upgrade of the formula (\ref{smalldelta'}) in the regime of parameters indicated by the Table 1. 
Figure \ref{fig:ky} gives a comparison between the theoretical growth rate predicted by Eqs. (\ref{dispPorcelli}), (\ref{smalldelta'}) and (\ref{corrected}), and the numerical growth rate $\gamma_N$ as a function of the wave number $k_y$. According to these tests, $\gamma_u$ seems to give a very good prediction for wave numbers $k_y > 1.1$. The discrepancy observed for lower values of $k_y$ comes from the fact that the condition allowing the use of the \textit{constant $\psi$ approximation}, $(\gamma d_e/ (k_y \rho_s)) \Delta' \ll 1$, is no longer satisfied for a small wave number, and for $\Delta' > \rs/(2 d_e^2)$, the solution is no longer real. 

\newpage


\subsection{Numerical results for $\bee \ne 0$}
We now proceed to a numerical study of the model (\ref{conteiso}) and (\ref{momeiso}), complemented by (\ref{qncondiso}), (\ref{ampparcondiso}) and (\ref{ampperpcondiso}).  This will allow to take into account the effects of finite $\beta_e$.

\begin{figure}
    \centering
\includegraphics[trim=0 0 0 0, scale=0.8]{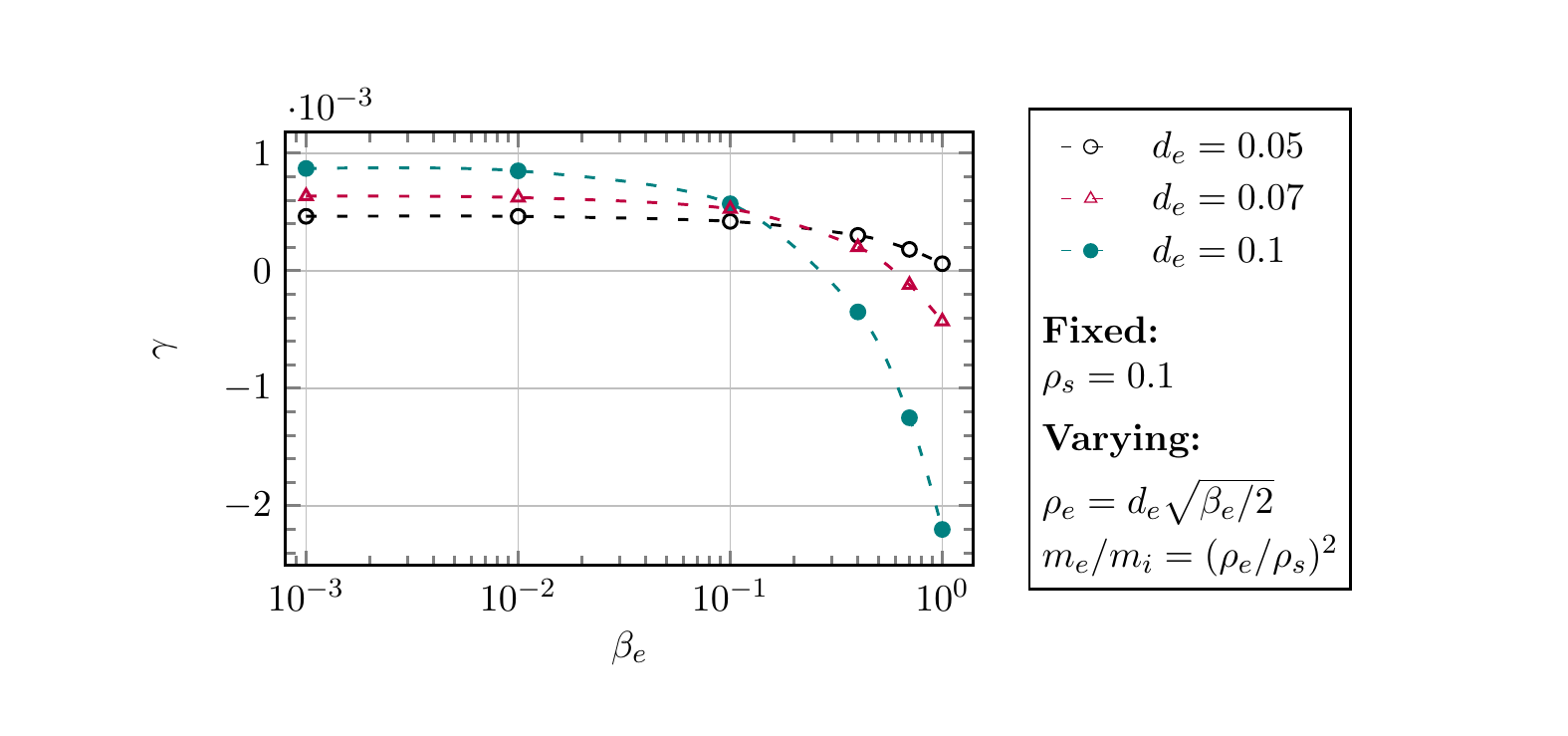}
     \caption{Numerical growth rates of the collisionless tearing mode as function of $\beta_e$, for three different values of $d_e$. The box length along $y$ is such that $-0.45\pi<y<0.45 \pi$, yielding a value of the tearing instability parameter of $\Delta'=0.067$ for the largest mode in the system. We stand in a very small $\Delta'$ regime, close to a marginal stability when $\bee < 0.1$. One sees that for higher values of $\bee$, and depending on the value of $d_e$, the mode is stabilized. }
    \label{fig:lin1} 
\end{figure}

The numerical set-ups are the same as those presented in the previous Section, relative to the equilibrium (\ref{equilibrium}), but the code accounts now for finite $\beta_e$ effects. The gyroaverage operators are introduced as they are defined in the Fourier space by Eqs. (\ref{op1}) and (\ref{op2}). 
For the linear tests we focus on a weakly unstable regime for which $0<\Delta'<1$. The strongly unstable case shows interesting behaviors in the non-linear phase and will be studied in the next Section. For all the tests, we will use $\lambda=1$.
In order to isolate the contribution coming from purely varying $\beta_e$, we first scan $ \bee $ from $ 10^{-3}$ to $1$ while $\rho_s$ and $d_e$ remain fixed, which is equivalent to considering a different mass ratio for each $\bee$ value.
We recall that the parameters are indeed linked by the relations
\begin{align} \label{rhoerel}
&   \rho_e = \rho_s \sqrt{\frac{m_e}{m_i}} = d_e \sqrt{\frac{\bee}{2}}.
\end{align}
Then we repeat the scan for different values of $d_e$. The results are presented in Fig. \ref{fig:lin1} and show that the single effect of $\bee$ in the model equations is stabilizing the tearing mode. This is consistent with the results obtained in the gyrokinetic and non-collisional study of \cite{Num11}, where $\bee$ and the mass ratio are also varied. 
Figure \ref{fig:lin1} also shows the competition between the destabilizing effect of the electron inertia and the stabilizing effect of $\bee$. For this set of parameter, the influence of $\bee$ on the weakly unstable regimes is almost negligible until $\bee=1$. For relatively low values of $\bee$, the highest growth rate corresponds to that for which the parameter $d_e$ is the largest. We recall in fact, from Section 3.1, that, for $\beta_e \ll 1$, the formulas (\ref{smalldelta'}) and (\ref{corrected}) hold. Such formulas, for $d_e \ll 1$, predict that the growth rate increases linearly with $d_e$. Conversely, when $\bee$ becomes large enough, as appears for $\bee > 0.15$, the growth rate for
which $d_e$ is the largest, decreases drastically under the effect of the finite $\rho_e$ and of the parallel magnetic perturbations induced by $\beta_e$.

Some information about the stabilizing role of $\bee$ can be inferred by taking the small FLR limit of the equation (\ref{momeiso}), which consists in considering the regime of parameters
\begin{align}  \label{reg1}
&d_e \ll 1 , \qquad \rs \ll 1 , \qquad \frac{d_e}{\rs} \ll 1, \qquad \bee = O(1),
\end{align}
and assuming,
\begin{equation}   \label{reg2}
\lapp = O(1).
\end{equation}
If we retain the first-order FLR corrections as $d_e ,\rs \rightarrow 0$, the resulting Ohm's law reads
\begin{align}
&\frac{\pa}{\pa t}\left( \apar +\left(\frac{\bee}{4}-1\right) d_e^2 \lapp \apar \right)+\left[ \phi  ,   \apar +\left(\frac{\bee}{4}-1\right) d_e^2 \lapp \apar  \right] \nno \\
&+\left( \frac{\bee}{4} d_e^2  + \rs^2\left(\frac{\bee}{2 + \bee} -1\right)\right)[\lapp \phi , \apar]=0.  \label{mom1flr}
\end{align}
The new contributions in Eq. (\ref{mom1flr}) are those due to finite $\bee$ and are not present in the usual two-field model by \cite{Sch94}. In particular, the contributions proportional to $(\bee/4)d_e^2$ come from electron FLR effects and the contribution proportional to $\bee\rs^2/(2+\bee)$ is due to the presence of the finite $\bpar$. In Eq. (\ref{mom1flr}), comparing with Eqs. (\ref{fluid1})-(\ref{fluid2}), it is possible to identify an effective electron skin depth $d_e'$ and an effective sonic Larmor radius $\rs'$, given by, 
\beq
d_e' = \sqrt{1 - \frac{\bee}{4}}d_e,
\eeq
and 
\beq
\rs ' = \sqrt{ \rs^2 \left(1 - \frac{\bee}{\bee +2}\right) - d_e^2\frac{\bee}{4}},
\eeq
respectively. Therefore, considering, as first approximation, the relation \ref{smalldelta'} with $d_e'$ and $\rho_s'$ replacing $d_e$ and $\rho_s$, respectively, one can identify some of the stabilizing effects of $\beta_e$, given that $d_e' < d_e$ and $\rs' < \rs$. 
However, the small FLR limit (\ref{reg1})-(\ref{reg2}) only gives us a limited insight, as it neglects higher-order derivatives contributions coming from the gyroaverage operators, which can become important around the resonant surface. On the other hand, this insight is arguably easier to obtain with the gyrofluid model, with respect to the gyrokinetic model.

\begin{figure}
    \centering
\includegraphics[trim=0 0 0 0 , scale=0.64]{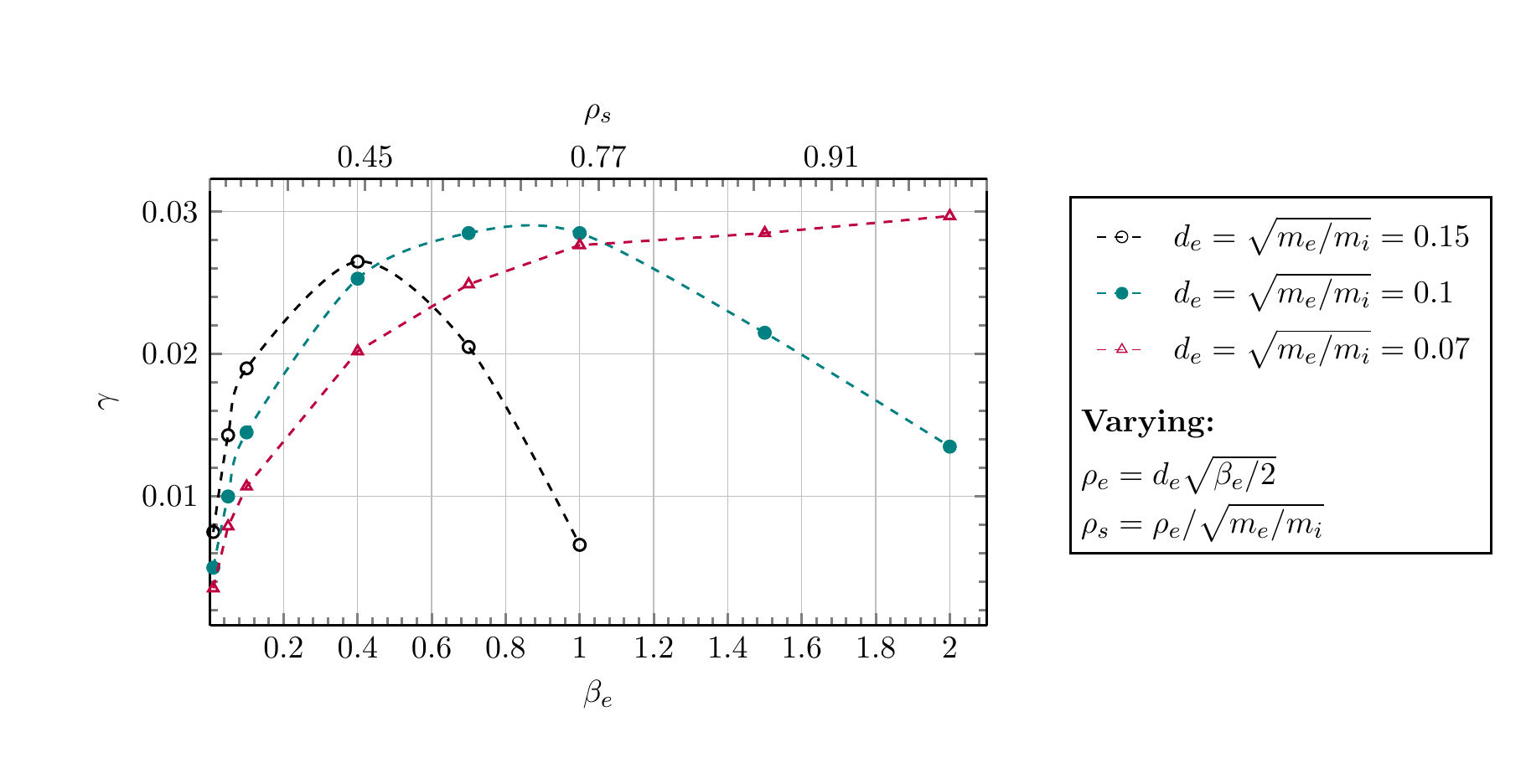}
    \caption{Numerical growth rates of the collisionless tearing mode as function of $\beta_e$ and $\rs$, for different values of $d_e= \sqrt{m_e/m_i}$. The box size is $- \pi < x < \pi$, $- 0.47\pi< y < 0.47\pi$, which leads to $\Delta'=0.59$. }
    \label{fig:lin2}
\end{figure}

\begin{figure}
    \centering
\includegraphics[trim=0 0 0 0, scale=0.8]{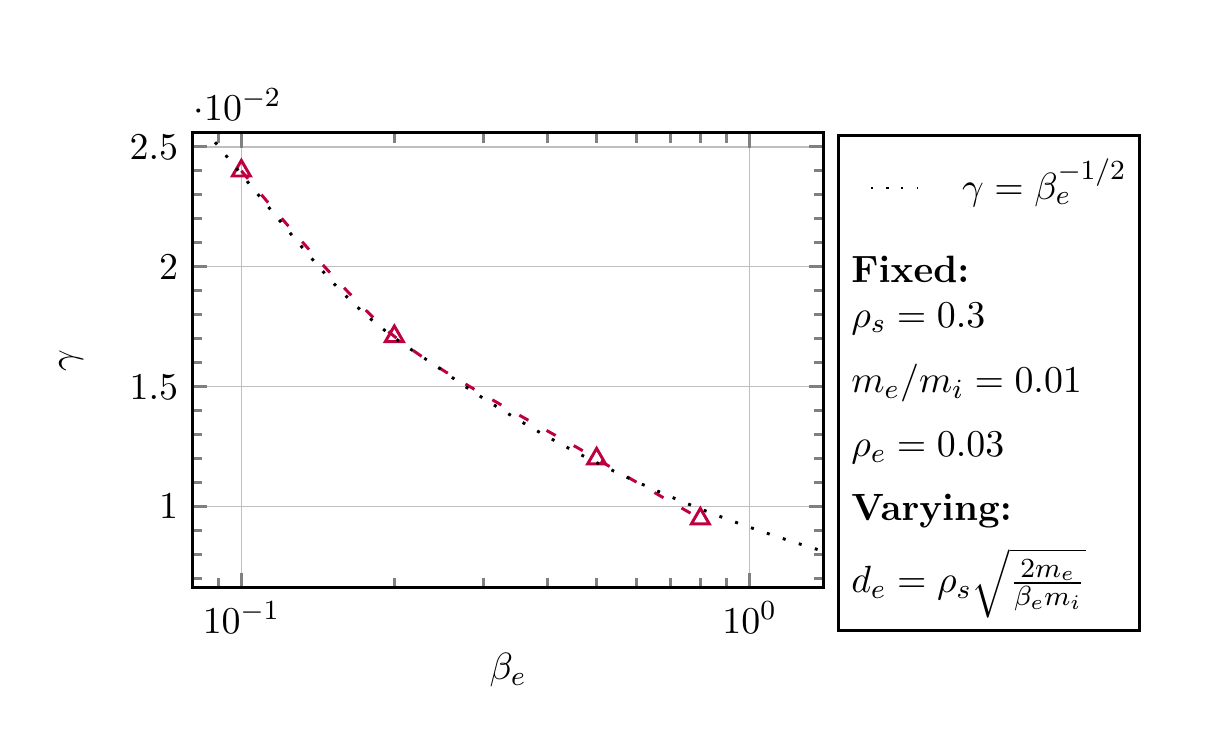}
     \caption{The value of $d_e$ for each run increases as $d_e=\sqrt{2 m_e /(\bee m_i)}\rs $. The box size is $- \pi < x <\pi$, $- 0.47\pi< y < 0.47\pi$. The numerical values (triangles) are compared with the curve $\gamma=\beta_e^{-1/2}$ (dotted line), which is the scaling predicted by \cite{Fit07} on the basis of a fluid model, and confirmed by gyrokinetic simulations by \cite{Num11}. The comparison shows that also our gyrofluid model confirms such scaling.}
    \label{fig:lin3} 
\end{figure}

A further analysis we carried out consists of investigating the effect of $\beta_e$ on the linear growth rate, but at a fixed mass ratio. Physically, this might be interpreted as investigating the effect of the variation of the equilibrium electron temperature $T_{0e}$ or of the background density $n_0$ of the plasma, on the stability of the tearing mode. In order to keep a constant mass ratio during the scan in $\bee$, we carried out a study with $\bee$ ranging from $10^{-3}$ to $2$ with $\rho_s$ varying simultaneously.
We fix the relation $ d_e = \sqrt{m_e/m_i} $ (implying $\rho_s=\sqrt{\beta_e / 2}$) and we evaluate the cases $ d_e = $ 0.07, $ d_e = $ 0.15, $ d_e = $ 0.1.
Figure \ref{fig:lin2} shows that when $\bee$ and $\rho_s$ are increased simultaneously there seems to be a competition between the destabilizing effect of $\rho_s$ and the stabilizing effect of $\bee$. Also in this case, the behavior at small $\beta_e$, can be interpreted on the basis of the formulas (\ref{smalldelta'}) and (\ref{corrected}), predicting an increase of the growth rate with increasing $\rho_s$. When electron FLR effects come into play at larger $\beta_e$, the growth rates decreases. The values chosen for the mass ratio are not realistic but make it possible to reduce the need of grid points. In the case of the artificial value of $d_e= \sqrt{m_e/m_i} = 0.15$, the stabilizing effect takes over the destabilizing effect of $\rho_s$ even for $\beta_e <1$. However, for the case $\sqrt{m_e/m_i} = 0.07$, much closer to a real mass ratio, the effect of $\rho_s$ appears to be dominant. Indeed, decreasing $d_e$ at a fixed $\beta_e$ amounts to decreasing $\rho_e$. Thus, for $d_e =0.07$ the stabilizing effect of the electron FLR terms gets weakened, with respect to the other values of $d_e$, even at large $\beta_e$.
\\
Figure \ref{fig:lin3} shows the variation of the growth rate of the tearing instability as a function of $\bee$, for a fixed value of $\rho_s = 10 \rho_e = 0.3$. The obtained results are confirming the scaling of the growth rate as $\beta_e^{-1/2}$ (or, equivalently, as $d_e$)  has been determined with the gyrokinetic study of \cite{Num15}. This shows the capability of the gyrofluid model to reasonably reproduce gyrokinetic results (\cite{Num11, Num15}) and the fluid theory of \cite{Fit07}, in a quantitative way.

\subsubsection{Hot ion limit, $\tau_i \rightarrow +\infty$}
\begin{figure}
    \centering
\includegraphics[trim=0 0 0 0 , scale=0.6]{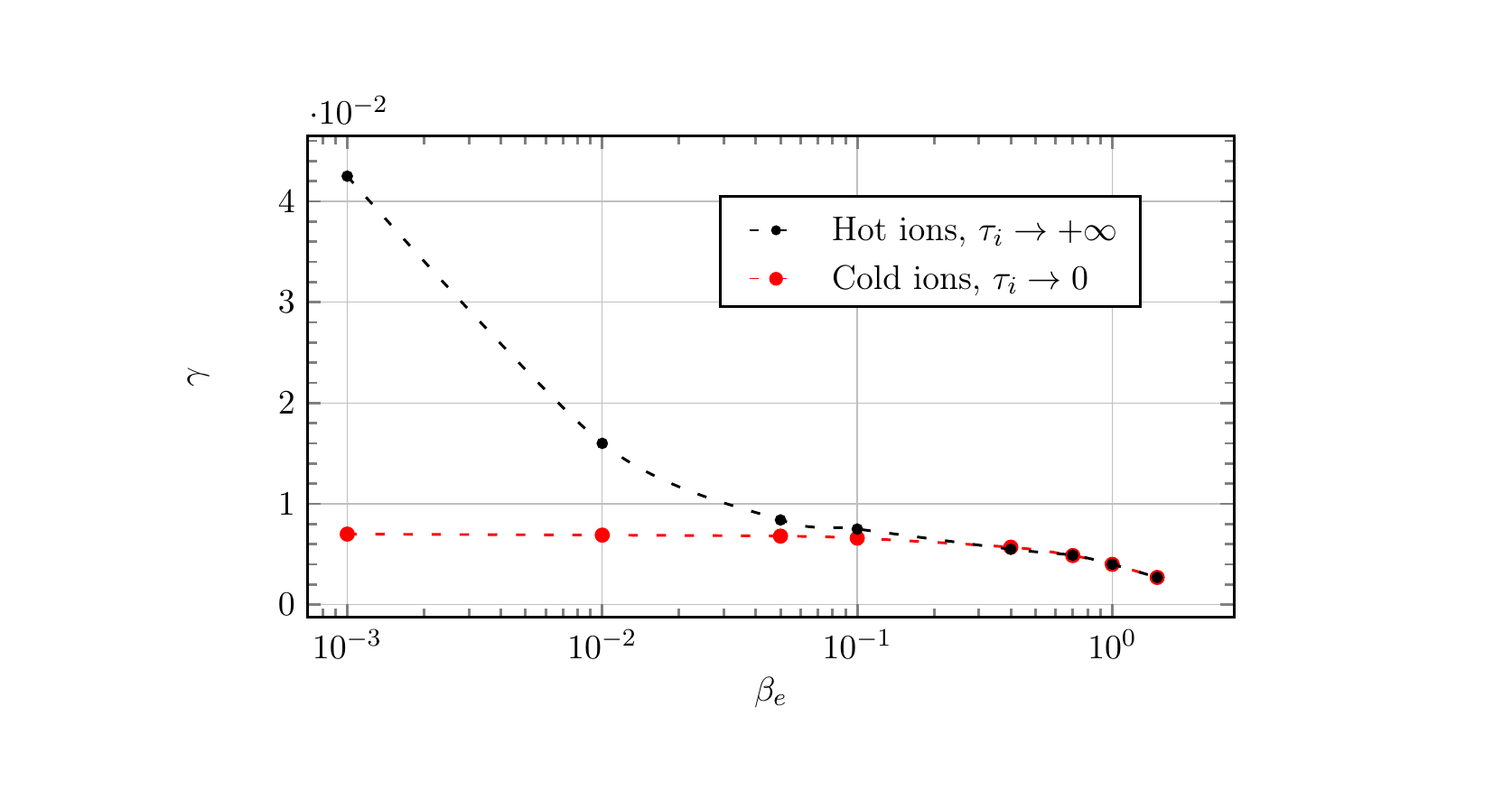}
    \caption{Comparison between the linear growth rate obtained in the cold-ion regime and the hot-ion regime. The box size is $- \pi < x <\pi$, $- 0.47\pi< y < 0.47\pi$, which leads to $\Delta'=0.59$. }
    \label{fig:hotion}
\end{figure}

In this article we have focused, so far, on the cold ion limit, but in this Subsection we temporarily deviate from the cold-ion case, to consider the opposite limit, in which $\tau_i=\tau_{\perp i} \rightarrow +\infty$. The sole purpose of this Subsection is to have a consistent and concise comparison of the two regimes, therefore we will only study the linear behavior of the hot ion limit and leave the study of its non-linear evolution for a future work.
The hot-ion limit can actually be of greater interest for space plasmas such as the solar wind. The ion gyrocenter density fluctuation and the ion gyrocenter parallel velocity are still neglected, and therefore the evolution equations remain unchanged. Only the assumption (\ref{cond2}) is taken in the opposed limit, which has an impact on the development of ion gyroaverage operators. The static relations (\ref{qncondiso}) and (\ref{ampperpcondiso}) are thus changed to
 \begin{align}
& \phi =  \frac{\rs^2 N_e}{\left(1 - \frac{\beta_e}{2}\right)G_{10e}-G_{10e}^{-1} }, \label{hotion1} \\ 
&  \bpar = \frac{\bee}{2 \rs^2} \phi. \label{hotion2} 
\end{align}
The linear results obtained in the hot-ion limit are compared to the results obtained in the cold-ion regime on Figure \ref{fig:hotion}. The parameters are $d_e=0.1$, $\rho_s=0.1$. 
Our results seem to indicate that, for $\bee> 0.5$, the growth rate is very insensitive to the temperature of the ions, which is in agreement with the results obtained by \cite{Num11}.
Studies have been carried out with arbitrary ratio between the equilibrium ion and electron temperature in the low-$\beta$ limit, by \cite{Por91, Gra99}, and predict that the growth rate is significantly higher when the temperature of the ion background temperature is higher than that of the electrons. This is indeed what we observe for $\bee<10^{-2}$.

\section{Nonlinear phase}\label{sec:nonlin}

\begin{figure}
    \centering
\includegraphics[trim=0 0 0 0 , scale=0.35]{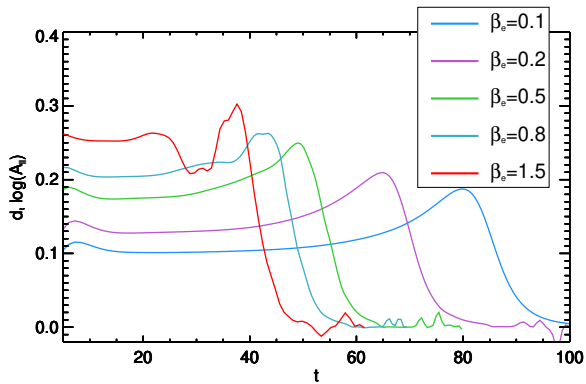}
    \caption{Plot of the effective growth rate $\frac{d}{dt} \log\left| \apar^{(1)} \left(\frac{\pi}{2},0,t \right)  \right|$, as a function of time. The corresponding values of $\beta_e$ are  shown in the table. The value of the electron skin depth is kept fixed to $d_e=0.08$, whereas $\rho_s$ is varied (and ranges from $0.17$ to $0.69$) so to keep the mass ratio fixed to $m_e/m_i=0.01$. All the growth rates, except for the case $\beta_e=1.5$ exhibit the same behavior, characterized by linear, faster than exponential and saturation phase. The case $\beta_e=1.5$ exhibits also a slow-down phase.  }
    \label{fig:nonlin1}
\end{figure}
To study the impact of a finite $\bee$ on the non-linear evolution of the magnetic island, we focus on the strongly unstable case, $\Delta'=14.31$, resulting from a box length along $y$ given by  $- \pi< y < \pi$. In this case, the mode $m=2$ has a positive tearing parameter $\Delta'_{2} = 1.23$. The higher harmonics are linearly unstable. The box along $x$ is chosen to be $-1.5 \pi < x < 1.5 \pi$ and allows to reach a large island without incurring in boundary effects. 
We make use of a resolution up to $2880\times2880$ grid points.
The mass ratio will be taken as $m_e/m_i = 0.01$ for the following tests. \\
The first tests are carried out by making a scan in $\bee$ from $\bee = 0.1$ to $\bee=1.5$ while keeping $d_e = 0.08$ and varying $\rho_s$ as $\rho_s=0.8 \sqrt{\bee} /\sqrt{2}$. 
Increasing $\bee$ and $\rs$ simultaneously in this way, as stated in Sec. 3.2, amounts to varying the electron background temperature $T_{0e}$.
Figure \ref{fig:nonlin1} shows the evolution in time of the effective growth rate, given by Eq. (\ref{numericalgrowthrate}), for each simulation. 
In all these cases, with the exception of $\bee=1.5$, we identify three phases; (1) a linear phase during which the perturbation evolution scales as $\mbox{exp}(\gamma t)$, (2) a faster than exponential phase, which is delayed in the case $\bee=0.1$, given that the linear growth rate is smaller, with respect to the case $\bee=0.8$ for which the instability reaches the nonlinear phase faster (3) a saturation during which the growth rate drops to $0$. We point out that, the fact that the linear growth rate increases with increasing $\bee$ is related to the fact that $\rho_s$ is also increased for each run. As discussed in the previous Section, the isolated effect of an increasing $\bee$ in the equations actually implies a stabilization of the linear growth rate.
For the case $\bee = 1.5$, we observe an intermediate phase, during which the growth of the island is slowed down.
It is also visible for the case $\bee = 0.8 $ that the growth rate shows a slowing down at $ t=38$ when it seemed to have already entered the explosive phase. Similar evolution and double faster than exponential phase have been studied in \cite{Com13}, where a finite ion Larmor radius is considered. 
\begin{figure}
    \centering
\includegraphics[trim=50 210 50 40 , scale=0.4]{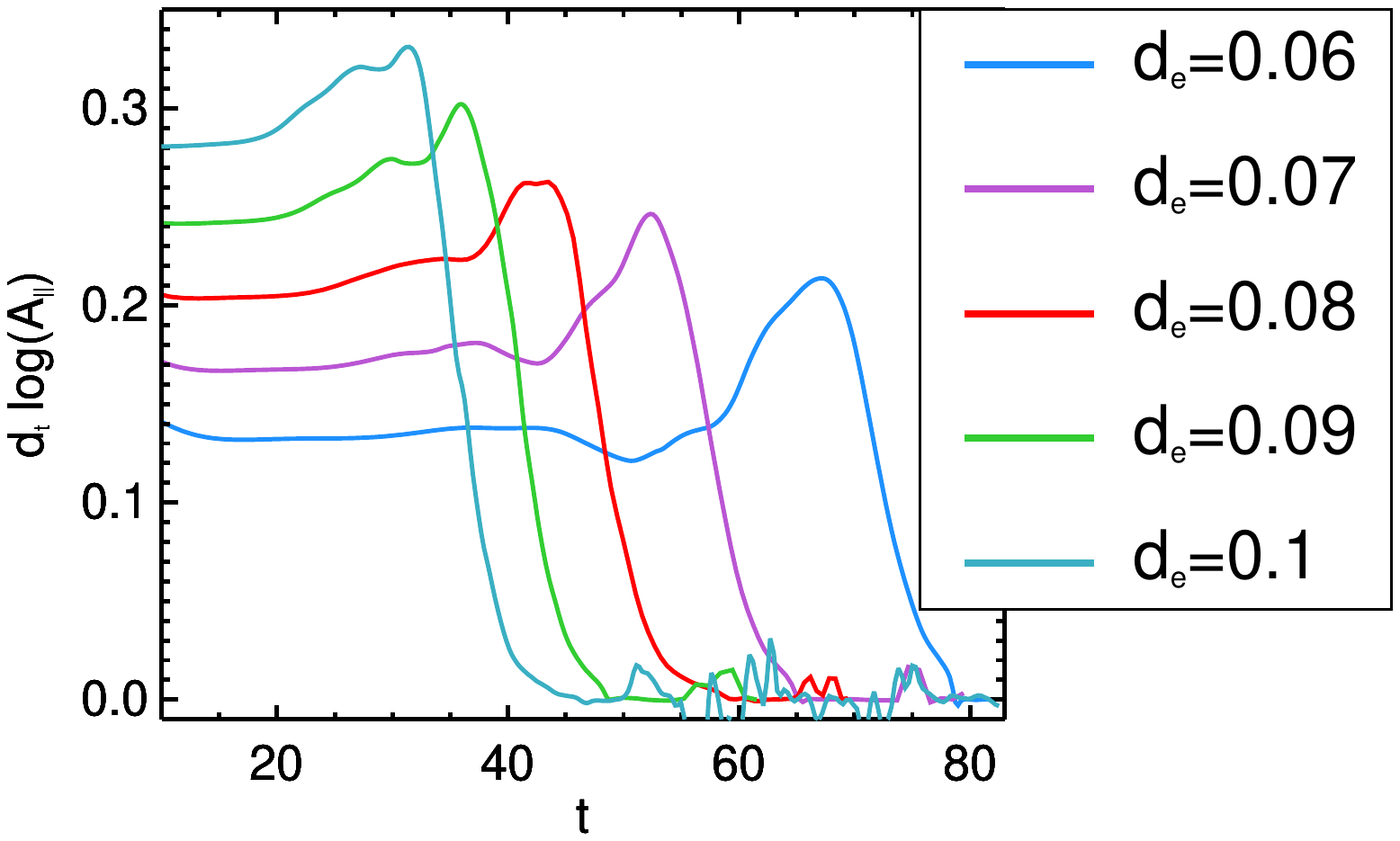}
\includegraphics[trim=20 0 50 0 , scale=0.75]{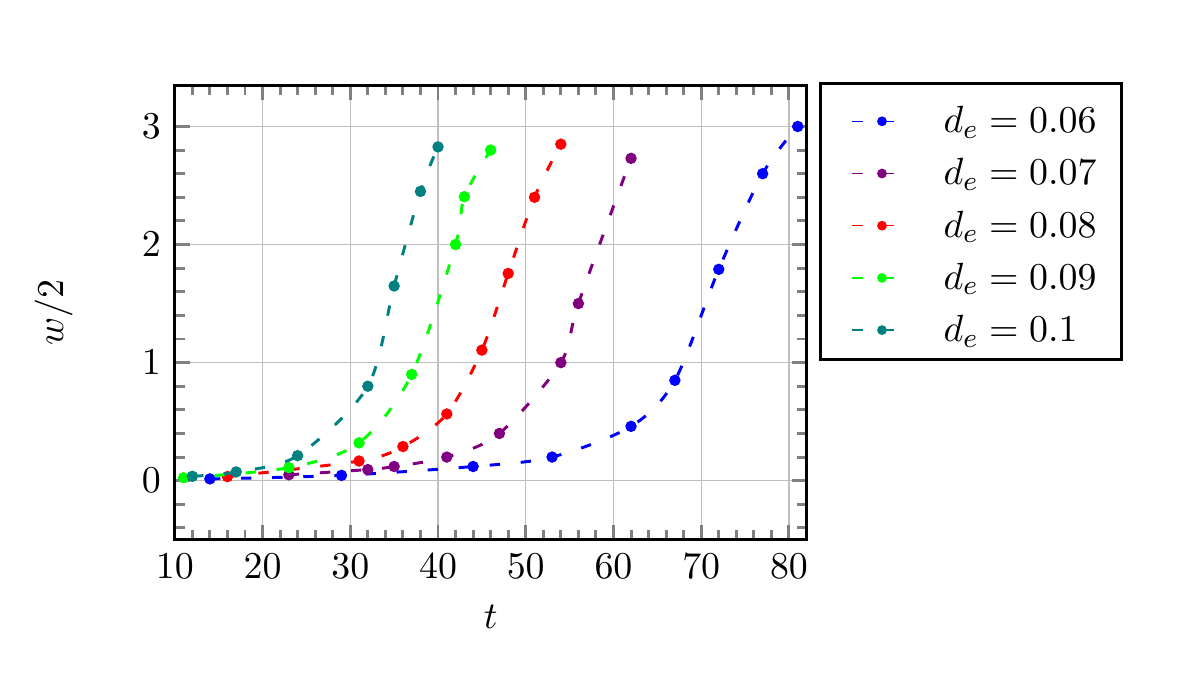}
    \caption{ On the left: plot of the effective growth rate $\frac{d}{dt} \log\left| \apar^{(1)} \left(\frac{\pi}{2},0,t \right)  \right|$, as a function of time. The parameters are $\bee =0.8$,  implying $\rho_e=\sqrt{0.4}d_e$ and $\rho_s =10 \sqrt{0.4}d_e$. On the right: Evolution of half-width of the magnetic island until saturation. The simulations correspond to those in the left panel.}
    \label{fig:nonlin2}
\end{figure}

We focus now on the case $\bee=0.8$. We scan the values of $d_e$ from $0.06$ to $0.1$, and $\rho_s =  10\rho_e = 10 \sqrt{0.4} d_e \approx 6.32 d_e$. The results are shown on Fig. \ref{fig:nonlin2}. These curves are compared for a fixed time unit (fixed $v_A$), while keeping $\bee$ and the mass ratio constant, which corresponds to varying $B_0 \sim n_0^{1/2}$ while keeping the electron temperature $T_{0e}$ fixed.  For the case of $d_e= 0.1$, which corresponds to $\rho_s \sim 0.63$, we observe that the slowdown  at the end of the linear phase. On the other hand, in the case of $d_e= 0.06$, for which $\rho_s = 0.37$, it appears at an earlier stage of the evolution process, when the nonlinear phase is already entered and it is followed by an explosive growth. The "double faster than exponential" behavior, which is observed in the cases $ d_e> 0.08 $, is similar to that observed in \cite{Com13} for large ion Larmor radius values.\\
The evolution of the width of the magnetic island for these five runs is shown on the right plot of Fig. \ref{fig:nonlin2}. The last point for each run corresponds to the width of the island when $\gamma_{max}$ is reached, just before the saturation phase. In conclusion, the growth of the island simply seems to be delayed, but the maximum width before saturation is identical for each case since the amount of initial magnetic energy is the same for each simulation.\\
\begin{figure}
    \centering
\includegraphics[trim=50 30 70 30 , scale=0.4]{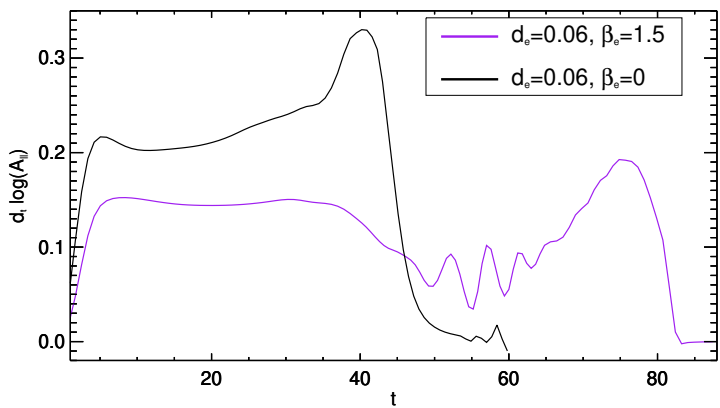}
    \caption{On the left: plot of the effective growth rate $\frac{d}{dt} \log\left| \apar^{(1)} \left(\frac{\pi}{2},0,t \right)  \right|$, for the cases $\bee=0$ (black curve) and $\bee=1.5$ (purple curve). The other parameters are $\rho_s = 0.519$ and $d_e=0.06$. On the right: Log of the time evolution of the reconnected flux at the X-point and the first 6 modes, from the simulation with $\bee=1.5$.}    \label{fig:nonlin4}
\end{figure}

The last test consists in studying an extreme case for which the slowing down phase is accentuated, which corresponds to the case of $d_e= 0.06$, $\rho_s = 0.519$, $\bee = 1.5$. 
We also perform the simulation for $\bee=0$, using a code that solves the fluid equations (\ref{fluid1}) - (\ref{fluid2}). Figure \ref{fig:nonlin4} shows the overplot of the evolution of the growth rate for both simulation as a function of time. 
The slowing down phase is followed by an oscillation of the non-linear growth rate. This oscillation was obtained in other tests for which $\bee = 1.5$ and is due to a slight displacement of the X-point during the reconnection. 
\subsection{Energy considerations}
\begin{figure}
    \centering
\includegraphics[trim=0 0 0 0, scale=0.28]{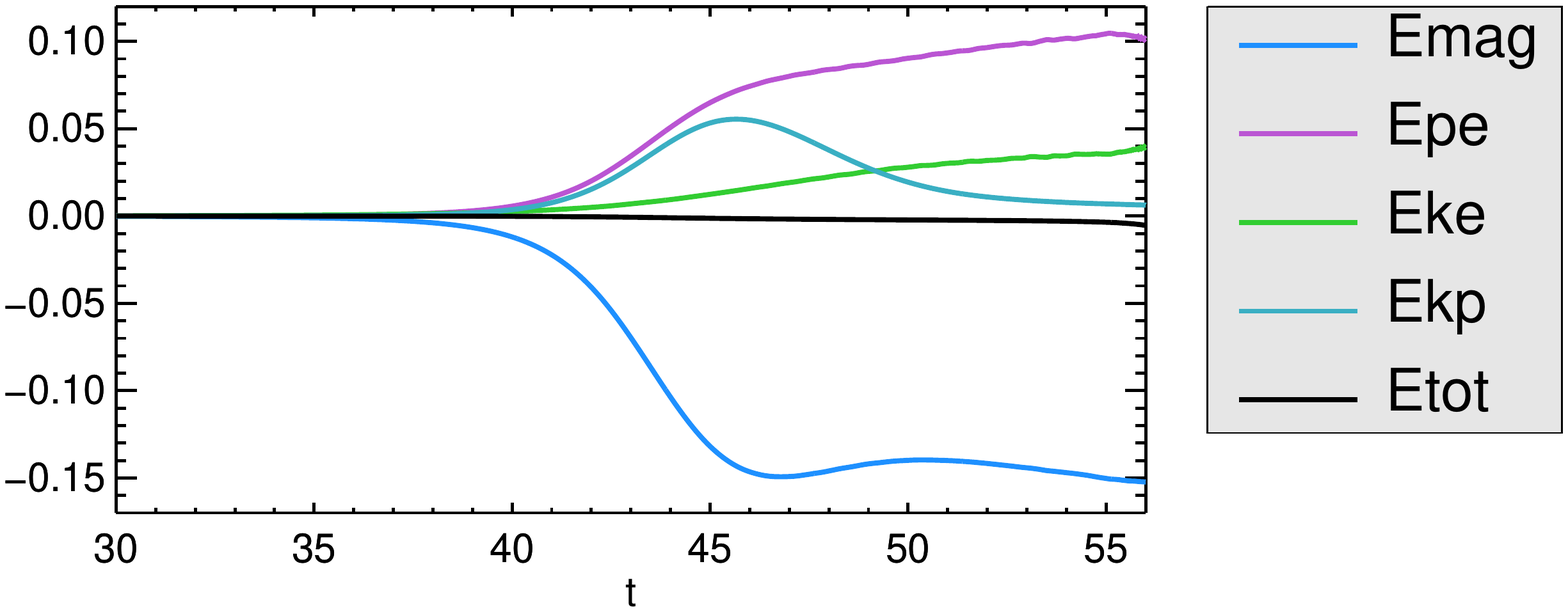}\vspace{-3cm}
\includegraphics[trim=0 0 0 0, scale=0.28]{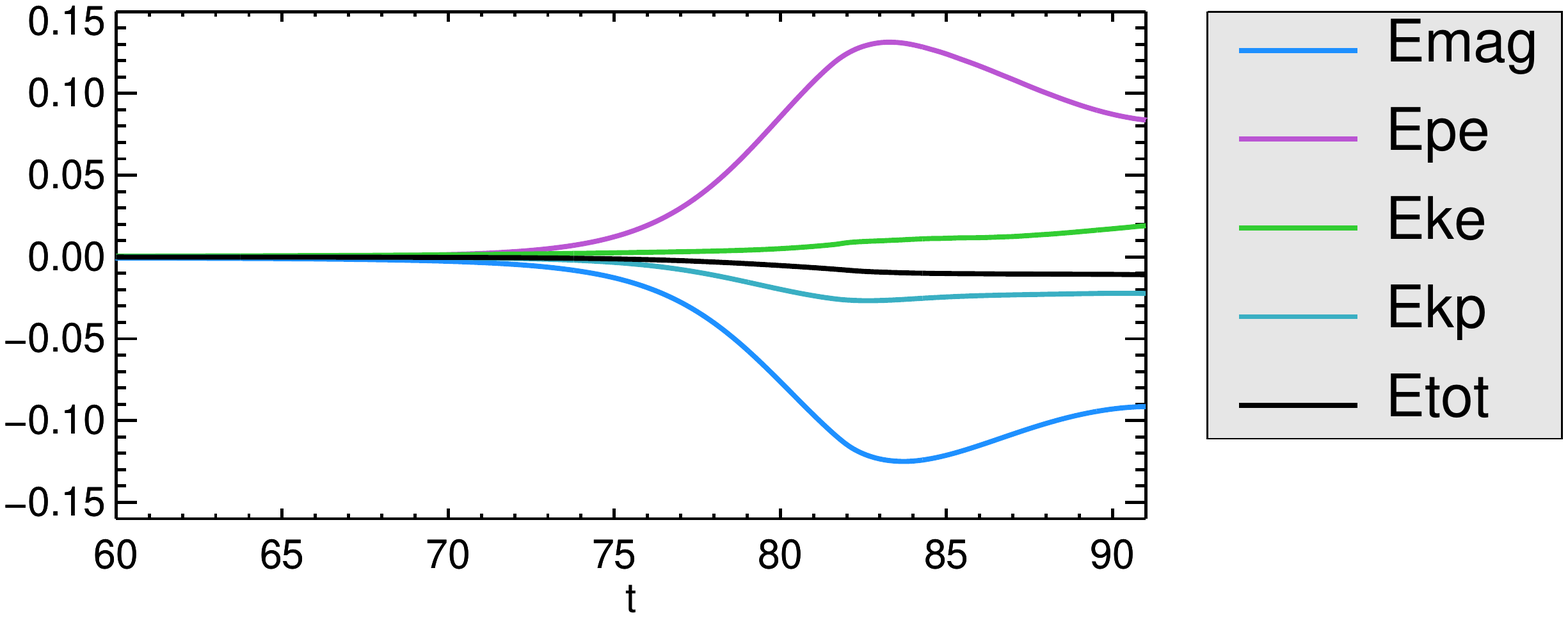}\vspace{-1cm}
    \caption{Time evolution of the energy variations for the cases $\bee=0$ (plot at the top) and $\bee=1.5$ (plot at the bottom). The parameters are $d_e=0.06$, $\rs=0.519$ and their corresponding growth rate is shown in Fig. \ref{fig:nonlin4}.}
    \label{fig:energy}
\end{figure}

The time variations of the different components of the energy for the cases $\bee=0$ and $\bee=1.5$, whose rate of growth is shown in Fig. \ref{fig:nonlin4}, are shown in Fig. \ref{fig:energy}. The variations are defined as $(1/2)\int dx^2( \xi(x,y,t) - \xi(x,y,0)) / H(0)$ where the function $\xi$ can be replaced by the different contributions of the Hamiltonian (\ref{ham2f}). 
In terms of the gyrofluid variables and in the presence of FLR effects, it is not obvious to identify the physical meaning of all the contributions to the energy. Therefore we use the terminology adopted in \cite{Tas18} and which refers to the fluid limit $\beta_e=0$.
The different contributions are, the magnetic energy, $E_{mag}$, for which $\xi= - U_e \gamue \apar $ (reduced to $|\nabla_{\perp} \apar|^2$ in the fluid case), the parallel electron kinetic energy, $E_{ke}$, for which $\xi = d_e^2 U_e^2$ (reduced to $d_e^2 ( \lapp \apar)^2$ in the fluid case), the energy due to the fluctuation of the electron density, $E_{pe}$, for which $\xi = \rho_s^2 N_e^2$ (reduced to $\rho_s^2 (\nabla_\perp^2 \phi)^2$ in the fluid case) and the perpendicular electrostatic energy of the electrons combined with the energy of the parallel magnetic perturbations, $E_{kp}$, for which $\xi = - ( \gamue \phi - \rho_s^2 2 \gde \bpar) N_e$ (reduced to $|\nabla_{\perp} \phi|^2$ in the fluid case).
We consider the simulation as being reliable until the time at which the percentage of the total energy that gets dissipated numerically (black curve) reaches $1 \%$.\\
By comparing the two simulations, one can see that there appears to be a comparable amount of magnetic energy being converted. The remarkable difference is the evolution of the component that combines the electrostatic energy and the energy of the parallel magnetic perturbations, $E_{kp}$, which, in the case $\bee=1.5$, also seems to be converted into electron thermal energy ($E_{pe}$), resulting in an increase in this component. This decrease of the electrostatic energy has been observed only in the case $\bee=1.5$. In the cases $\bee =0.8$, it appears that this component stays rather close to its initial value. \\
We also carried out the test with $\bee=1.5$ by artificially removing the parallel magnetic perturbation $\bpar$ from the code, and consequently it was not appearing in the expression of $E_{kp}$. It appeared first that the presence of $\bpar$ has a stabilizing effect on the tearing mode (which is consistent with the linear results discussed in Sec. 3.2), and secondly, the energy component $E_{kp}$ was slightly increasing instead of decreasing. This allows us to conclude that the energy related to the parallel magnetic perturbations is in fact the decreasing component that seems to be converted into electron thermal energy $E_{pe}$.\\ 
%


\section{Conservation laws of the model}\label{sec:inv}
In this Section we discuss the conservation laws of the gyrofluid model and its Lagrangian invariants.
Equations (\ref{conteiso})-(\ref{momeiso}) can be recast in the form
\beq  \label{lagr}
\frac{\pa A_\pm}{\pa t}+\mathbf{v}_\pm \cdot \nabla A_\pm=0,
\eeq
where 
\begin{align}
&A_\pm = \gamue \apar - d_e^2 U_e \pm d_e \rs N_e, \label{apm}\\
&\mathbf{v}_\pm = \hat{z} \times \nabla \left(\gamue \phi - \rs^2 2 \gamde \bpar \pm \frac{\rs}{d_e} \gamue \apar\right). \label{vpm}
\end{align}
We define by
\begin{align}
&\phi_\pm = \gamue \phi - \rs^2 2 \gamde \bpar \pm \frac{\rs}{d_e} \gamue \apar, \label{phipm}
\end{align}
the stream functions of the velocity fields $\mathbf{v}_\pm = \hat{z} \times \nabla \phi_\pm$.
The formulation (\ref{lagr}) makes it evident the presence of Lagrangian invariants, corresponding to the fields $A_\pm$,
in the model. Such Lagrangian invariants are advected by the incompressible velocity fields $\mathbf{v}_\pm$. The presence of such Lagrangian invariants is a feature common to many 2D Hamiltonian reduced gyrofluid models \citep{Wae09,WT_2012,Ker15,Tas19,Tas17,Pas18, Gra10, Gra15} and is related to the existence of infinite families of Casimir invariants of the Poisson bracket.

For Eqs. (\ref{conteiso})-(\ref{momeiso}) , such invariants correspond to the two families
\beq
C_+=\int d^2 x \, \mathcal{C}_+ (A_+), \qquad C_-=\int d^2 x \, \mathcal{C}_- (A_-),
\eeq
where $\mathcal{C}_\pm$ are arbitrary functions.
Equations (\ref{lagr}) imply that contour lines of the fields $A_\pm$ cannot reconnect, as the corresponding vector fields $\mathbf{B}_\pm=\nabla A_\pm \times \hat{z}$ are frozen in the velocity fields $\mathbf{v}_\pm$. On the other hand, the same model allows magnetic field lines to reconnect. In particular, it is useful to illustrate the mechanisms breaking the frozen-in condition in this model. This can be done by inspection of Eq. (\ref{momeiso}),  governing the evolution of $\apar$, and consequently, of the magnetic field in the plane perpendicular to the guide field, which is given by $\mathbf{B}_\perp=\nabla \apar \times \hat{z}$. Equation (\ref{mome}) can be rewritten in the following way:
\begin{align}
& \frac{\pa \apar}{\pa t} + \bu \cdot \nabla \apar \nno \\
&= -\frac{\mathcal{D}}{\mathcal{D} t} \left(\left( \frac{\bee}{4} -1\right) d_e^2 \lapp \apar + \sum_{n=2}^{+\infty} \left( \frac{\bee}{4n}-(-1)^{n-1}\right)\left(\frac{\bee}{4}\right)^{n-1}\frac{(d_e^2 \lapp)^n}{(n-1) !} \apar \right) \label{ohm} \\
&-\rs^2 \sum_{n=1}^{+\infty} \frac{1}{n!}\left( \frac{\bee}{4} d_e^2\right)^n [ {(\lapp)}^n \apar , N_e], \nno
\end{align}
where
\beq
\bu=\hat{z}\times\nabla(\gamue \phi - \rs^2 2 \gamde \bpar - \rs^2 N_e), 
\eeq
and where the operator $\mathcal{D}/\mathcal{D} t$ is defined by
\beq
\frac{\mathcal{D}f}{\mathcal{D} t}=\frac{\pa f}{\pa t}+[\gamue \phi - \rs^2 2 \gamde \bpar ,f]
\eeq
for a function $f$. In Eq. (\ref{ohm}) we also used the formal expansions 
\beq
\gamue=\sum_{n=0}^{+\infty}\frac{1}{n!} \left(\frac{\bee}{4} d_e^2 \lapp \right)^n, \qquad \gamue^{-1}=\sum_{n=0}^{+\infty}\frac{(-1)^n}{n!} \left(\frac{\bee}{4} d_e^2 \lapp \right)^n.
\eeq
The right-hand side of Eq. (\ref{ohm}) contains all the terms that break the frozen-in condition. Indeed, if the right-hand side of Eq. (\ref{ohm}) vanishes, the perpendicular magnetic field is frozen in the velocity field $\bu$. From Eq. (\ref{ohm}) one thus sees that the frozen-in condition can be violated by electron inertia (associated with the parameter $d_e$) and by electron FLR effects (associated with the combination $(\bee /4) d_e^2$). In the limit $\bee=0$ only electron inertia remains to break the frozen-in condition. On the other hand, because electron FLR terms are associated with the product between $\bee/4$ and $d_e^2$, in the limit $d_e=0$ both electron inertia and electron FLR terms disappear and the right-hand side of Eq. (\ref{ohm}) vanishes, thus restoring the frozen-in condition. We remark that the presence of a finite $\bee$ is also responsible for finite parallel magnetic perturbations $\bpar$. However, these do not violate the frozen-in condition for the perpendicular magnetic field, as they only contribute to modify the advecting velocity field $\bu$ (the parallel magnetic field lines, on the other hand, might undergo reconnection).

\begin{figure}
    \centering
\includegraphics[trim=80 0 30 0 , scale=0.22]{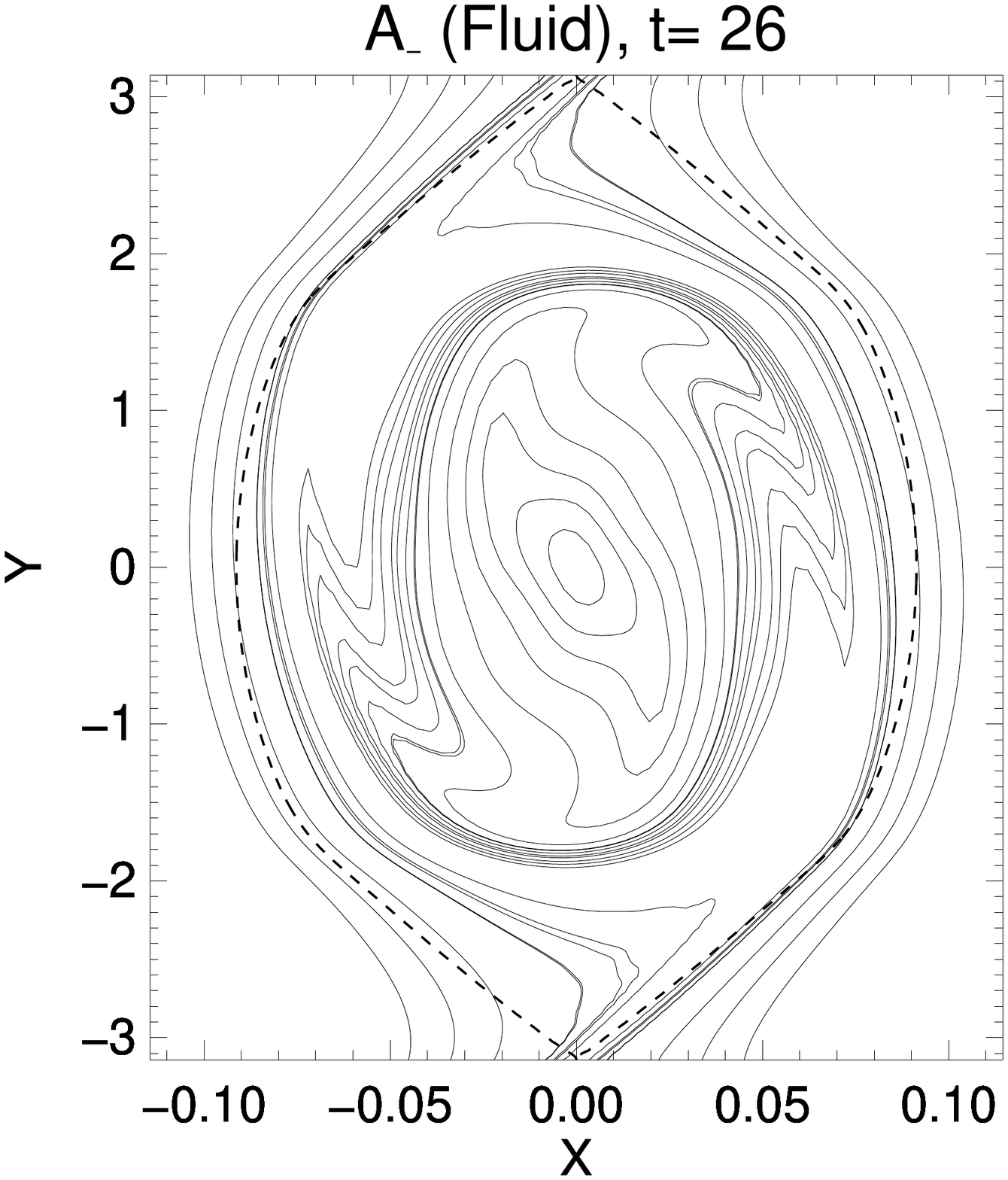}
 \includegraphics[trim=80 0 30 0 , scale=0.22]{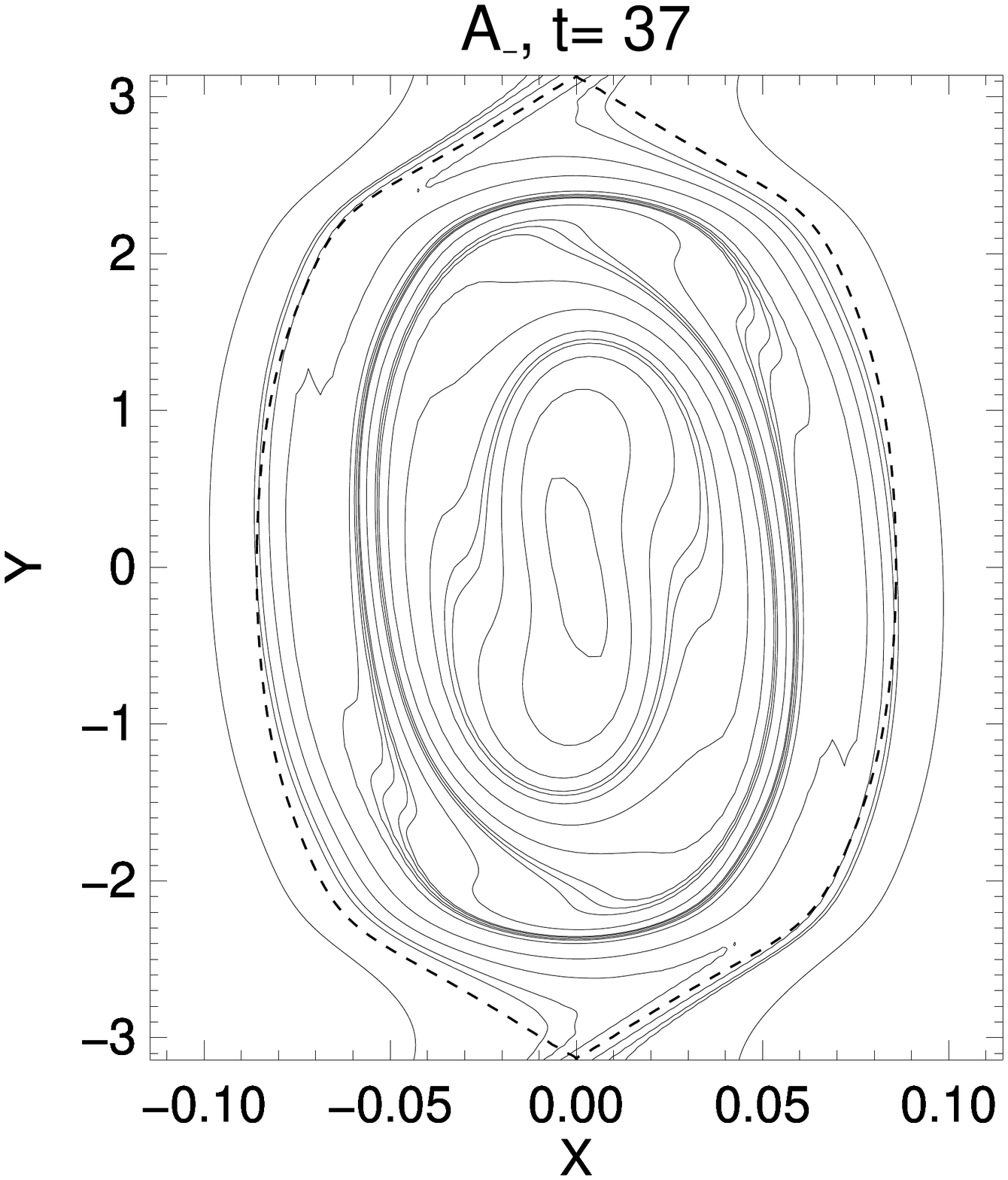} 
    \caption{
    Contour plot of the Lagrangian invariant $A_{-}$. Left panel: $\bee=0$, right panel: $\bee=1.5$. The parameters are $d_e=0.06$, $\rs=0.519$. The dashed lines are the separatrices. The contour plots refer to the normalized time $\gamma t= 5.18$}
    \label{fig:am}
\end{figure}

We consider here the qualitative structures of the contour plots of the Lagrangian invariants $A_\pm$ referring to the choice of parameters already adopted for Fig. \ref{fig:nonlin4}. From comparing the contour plots of $A_{-}$, in the case $\bee=0$ (left panel of Fig. \ref{fig:am}) and $\bee=1.5$ (middle panel of Fig. \ref{fig:am}), the structures look qualitatively similar. 
The contour lines of $A_{-}$ are induced by the velocity fields $\phi_{-}$ and undergo a phase mixing (the field  $A_{+}$ is winding up identically in the opposite direction, induced by $\phi_{+}$).
The duration of the transient and linear phases are not identical, consequently we compared the fields at the normalized time $\gamma t= 5.18$, which makes it possible to compare the fields when the islands are of comparable size so that they reached the same stage of evolution. The separatrices are displayed on each plot by dashed lines. We observe a different shape of the island in the two cases, which reflects the different distribution of the spectral power of the magnetic field. The effect of $\bee$ gives a more elongated island along $y$ and thinner along $x$. 
If we take a $\bee>1$ and keep a low enough mass ratio, then we are forced to stand in a regime with $\rs/d_e$  much greater than $1$. The ratio considered in this simulation is $\rho_s/d_e=8.65$. In this case $A_\pm$ is advected by a velocity field which can be approximated by $\mathbf{v}_\pm = \pm\hat{z} \times \nabla \left(  \frac{\rs}{d_e} \gamue \apar\right)$, since  $\phi_\pm$ tends to coincide with $\pm \frac{\rs}{d_e} \gamue \apar$. Performing other tests (whose results are not shown here) with $d_e \sim \rho_s$, $\beta_e \in \{0, 0.5\}$ and a mass ratio $20$ times higher, did not show any obvious difference in the mixing phase either. 
\begin{figure}
    \centering
\includegraphics[trim=0 0 0 0 , scale=0.2]{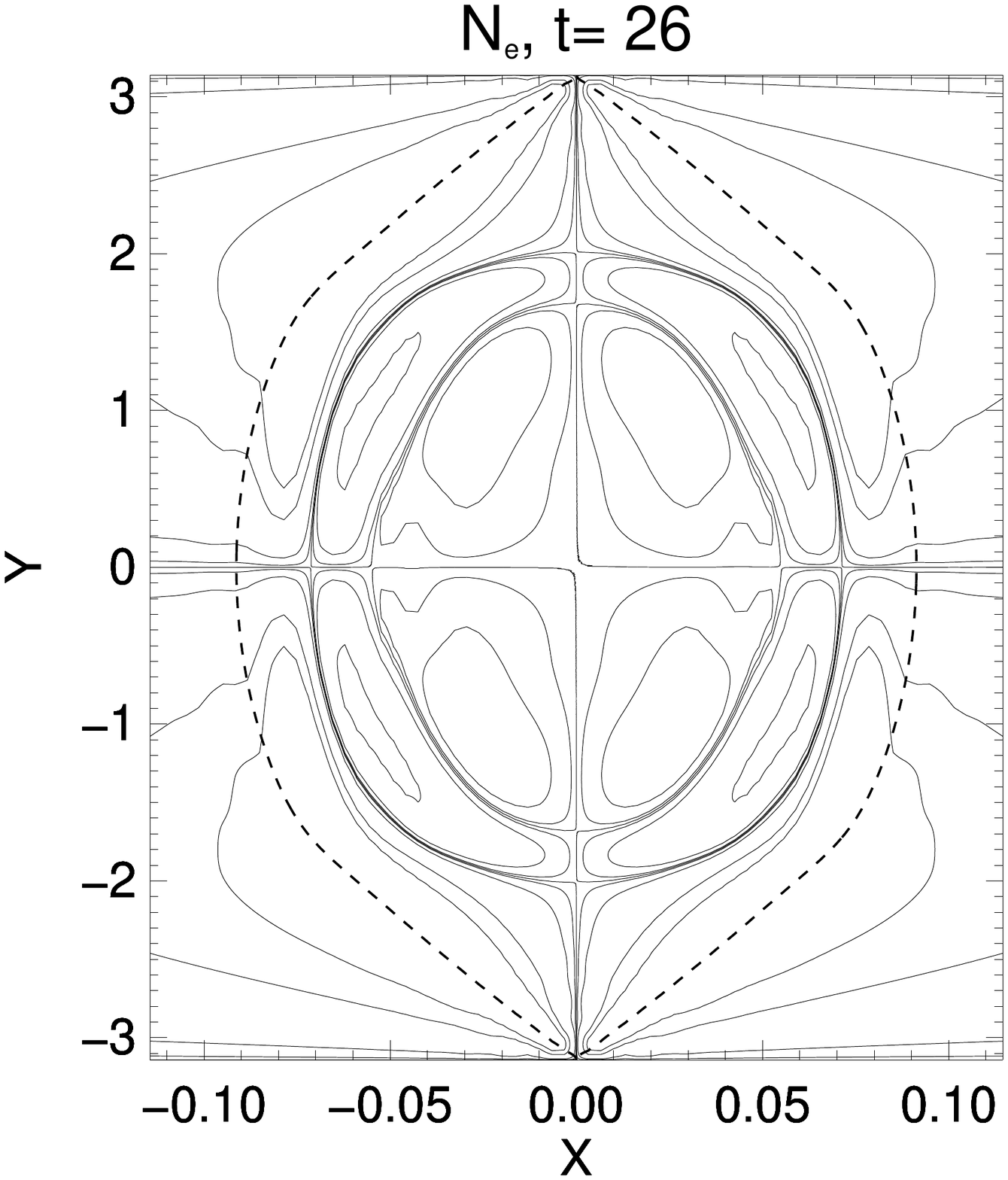}
 \includegraphics[trim=0 0 0 0 , scale=0.2]{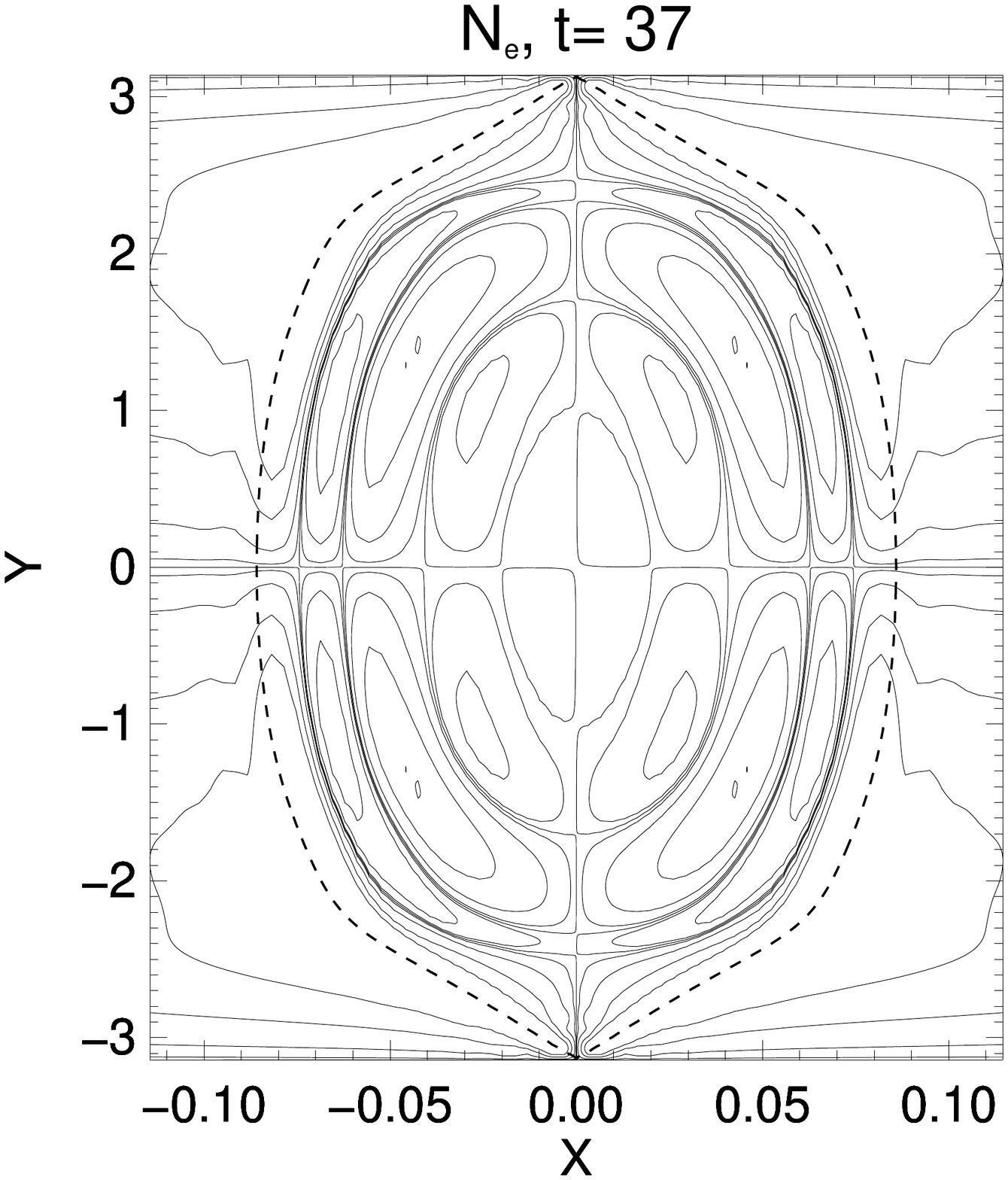} 
 \includegraphics[trim=0 0 0 0, scale=0.2]{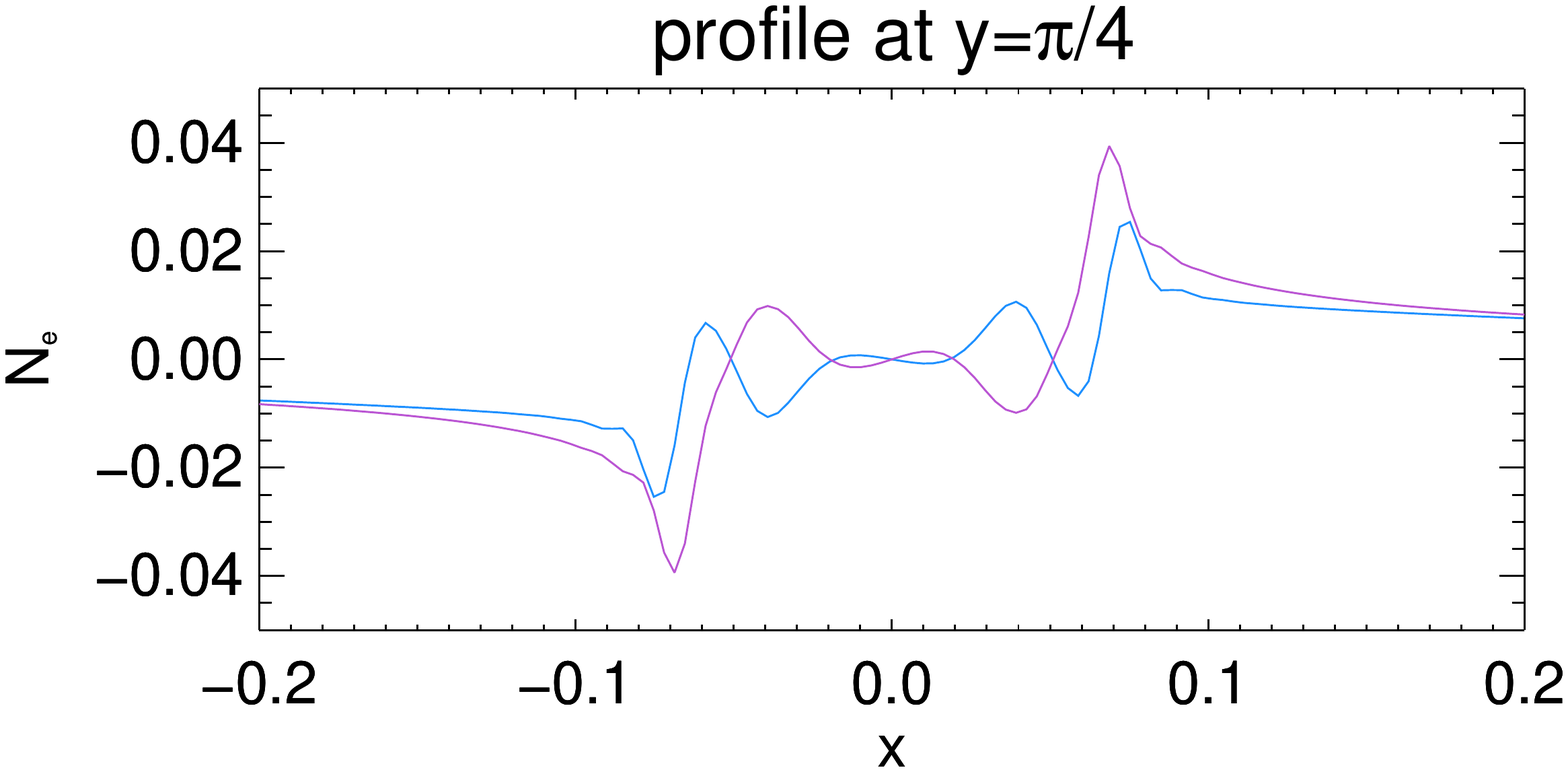} 
    \caption{
    Contour plot of the electron density. Left panel: $\bee=0$, middle panel: $\bee=1.5$. On the right panel are the profiles of $N_e$ at $y=\pi/3$ in the cases $\bee=0$ (purple) and $\bee=1.5$ (blue). The parameters are $d_e=0.06$, $\rs=0.519$. The dashed lines are the separatrices. The contour plots and profiles refer to the normalized time $\gamma t= 5.18$}
    \label{fig:ne}
\end{figure}

The electron density $N_e$ can be obtained by a linear combination of the invariants $A_\pm$
\beq
N_e = \frac{A_+ - A_-}{2 d_e \rs}.
\eeq
The contour plot of the electron density is displayed on Fig. \ref{fig:ne} and shows the fine structures produced by the mixing of the Lagrangian invariants $A_\pm$. The case $\bee=1.5$ shows nested quadripolar structures. 
From the difference between the profiles of $N_e$ on Fig. \ref{fig:ne}  it is visible that increasing $\bee$ will smooth the gradients in the inner region of the electron density. 

\section{Conclusion}

In this article, we have attempted to provide an overview of the impact of finite electron plasma beta effects on the tearing instability in non-collisional plasma with cold ions and a strong guide field. 
Adopting a gyrofluid model, we have studied the effects of electron gyration and of a parallel magnetic perturbation. 
There is a wide variety of systems for which this study can be useful, such as magnetosheat plasmas, where current sheets form in the presence of a guide field and a large $\bee$ value. Recently, for instance, studies of observations of the MMS space mission in the magnetotail have revealed electron-only reconnecting current sheet, where ions do not participate and where $\bee$ values can be observed to be greater than $1$ (\cite{Man20}).

Our main results are the following. First, increasing $\bee$ and $\rho_s$ while keeping $d_e$ and the mass ratio fixed, the evolution of the reconnection growth rate seems to be dominated by the destabilizing effect of $\rho_s$, up to a certain threshold where the effects of $\rho_e$ become important and the grfowth rate diminishes (Fig. \ref{fig:lin2}). This can also be interpreted as fixing the background density, $n_0$, the ion mass (so that $d_e$ is fixed) and the guide field amplitude $B_0$, while increasing the electron temperature $T_{0e}$. In the case of a small $\Delta'$ regime, a high $\bee$ can eventually stabilize the tearing mode and prevents reconnection from occurring. \\\
Secondly, in the nonlinear regime of the case $\rho_s \gg d_e$ with  $\beta_e \sim m_e/m_i \ll 1$, (which is referred to as being the fluid regime in this article), we retrieved the well-know collisionless faster than exponential growth which leads to an explosive growth of the magnetic island. However, when we increase $\bee$, this explosive paradigm is modified with the appearance of a slow down phase preceding the explosive growth.  \\
A further conclusion is that the effect of $\bee$ on the Lagrangian invariants of the gyrofluid model does not seem to reduce the filamentary structure, produced by a "phase mixing", characteristic of these invariants.

The results obtained with our gyrofluid model are in agreement with results obtained by gyrokinetic studies (\cite{Num11, Num15}). They also complement some two-fluid studies where a consistent accounting for $\beta_e$ effects, including both electron FLR and parallel magnetic perturbations were neglected (\cite{Sch94, Gra99, Del06, Fit07}).



\subsection*{Acknowledgements}
The authors acknowledge helpful discussions with Dimitri Laveder.
This work benefits from the support of the Ignitor project under the CNR contract DFM.AD003.261 (IGNITOR)-Del. CIPE n.79 del 07/08/2017.
The numerical simulations were performed using the EUROfusion high performance computer Marconi Fusion hosted at CINECA (project FUA35-FKMR) and the computing facilities provided by “Mesocentre SIGAMM” hosted by Observatoire de la Côte d’Azur.

\subsection*{Appendix: Calculation of $\gamma_u$}

We start from the linearized Eqs. (\ref{fluid1})-(\ref{fluid2}), using the equilibrium (\ref{equilibrium}) and the perturbations (\ref{perturbations}).
The perturbations are subject to the boundary conditions $\wapar , \,  \wphi \rightarrow 0$, as $x \rightarrow \pm  \infty$. We look for even solutions of $\wapar(x)$ and odd solutions for $\wphi(x)$, which are standard parities for the classical tearing problem. 

We consider the time variation of the perturbation being slow,
\beq
g=\frac{\gamma}{k_y} \ll 1,
\eeq
and the normalized electron skin depth as a small parameter, i.e.
\begin{equation}
d_e \ll 1.
\end{equation}
The linearized equations are given by
\beq \label{linear1}
\gamma ( \wphi'' - k_y^2 \wphi) - i k_y \apar \byo'' + i k_y \byo ( \wapar'' - k_y^2 \wapar)=0,
\eeq
\beq \label{linear2}
\gamma( \wapar - d_e^2 ( \wapar'' - k_y^2 \wapar)) + i k_y \wphi (\byo - d_e^2 \byo'') - i k_y \rho_s^2 \byo ( \wphi'' - k_y^2 \wphi ) =0, 
\eeq
where $\byo = - \d d \apar^{(0)} / \d d x $ is the equilibrium magnetic field. In order to solve (\ref{linear1}) and (\ref{linear2}) we have to adopt an asymptotic matching method because the vanishing of the two small parameters $g$ and $d_e$ leads to a boundary layer at the resonant surface $x=0$. We will consider two spatial regions involving two spatial scales. Far from the resonant surface, located at $x=0$, the plasma can be assumed to be ideal and electron inertia can be neglected. This region is commonly called the \textit{outer region}. Close to the resonant surface, we will proceed to a spatial rescaling and get to a scale at which electron inertia becomes important and drives the reconnection process. This second region is called the \textit{inner region}. We anticipate that we will find a second boundary layer inside the inner region and will need the use of a second asymptotic matching.


\subsection{Outer region}
As mentioned before, we assume $d_e \ll1$ and $g \ll 1$. We then neglect terms of order $d_e^2$ and $g^2$ in Eqs. (\ref{linear1}) and (\ref{linear2}). The outer equations are given by
\beq
\wapar''_{out} - \left( k_{y}^{2}  + \frac{\byo^{''}}{\byo} \right)\wapar_{out}  = 0 \label{outer1}
\eeq
\begin{equation}
    \widetilde{\phi}_{out}(x)=\frac{i g \wapar_{out}(x)}{B_{y0}}, \label{outer2}
\end{equation} 
where we indicate with the prime symbol, the derivative with respect to the argument of the function.
%
The solution for $\tilde{A}_{out}$ is given by
\begin{align} \label{outersolutionA2}
  &   \wapar_{out} (x)= e^{-\frac{\left| x\right|  \sqrt{ \lambda^2  k_y^2+4}}{ \lambda}}
   \left(\frac{15 \tanh ^3\left(\frac{\left| x\right|
   }{ \lambda}\right)}{ \lambda^2  k_y^2 \sqrt{ \lambda^2
    k_y^2+4}}+\frac{15 \tanh ^2\left(\frac{\left| x\right|
   }{ \lambda}\right)}{ \lambda^2  k_y^2} \right. \nno \\
& \left.+\frac{\left(6
   \left( \lambda^2  k_y^2+4\right)-9\right) \tanh
   \left(\frac{\left| x\right| }{ \lambda}\right)}{ \lambda^2
    k_y^2 \sqrt{ \lambda^2  k_y^2+4}}+1\right)
\end{align} 
From Eq. (\ref{outer2}), on the other hand, one sees that the solution for $\tilde{\phi}_{out}$ is not defined at the resonant surface $x=0$, where $B_{y0}$ vanishes. This indicates the presence of the above mentioned boundary layer at $x=0$. We measure the logarithmic derivative of the of the discontinuity of the outer solutions  (\ref{outersolutionA2}) at $x=0$ with the formula (\ref{Delta'}) of the standard tearing parameter and we obtain the expression
\begin{equation} \label{Delta'2}
    \Delta' =  \frac{2 \left(5- \lambda^2 k_y^2\right) \left(\lambda^2 k_y^2+3\right)} {\lambda^3 k_y^2 \sqrt{\lambda^2 k_y^2+4}}.
\end{equation}
In the limit $|x| \rightarrow 0 $ the solution for $\tilde{A}_{out}$ can be develop using its Taylor expansion
\beq \label{expansionA}
\wapar_{out}=  1 + \frac{\Delta'}{2} |x| + O(x^2).
\eeq
If $\Delta'$ is small enough, the solution $\wapar$  can be approximated to be equal to $1$ in the region where $x \ll 1$.
This is standard procedure called the \textit{constant $\psi$ approximation} (\cite{Fur63}).  

\subsection{Inner region: first boundary layer}

In the inner region, we proceed to a first spatial rescaling using an inner variable, $\hat{x}$, such that
\beq \label{changeofvariable}
x =\epsilon \hx,
\eeq
where $\epsilon \ll 1$ is a stretching parameter. The rescaling (\ref{changeofvariable}) implies  $k_y\ll \partial_{\hat{x}} $, and allows to use a Taylor expansion of the equilibria (\ref{equilibrium})
\beq
\byo(\epsilon \hat{x}) = \frac{2 \hx \epsilon}{\lambda} + O(\epsilon^2).
\eeq
We obtain the two inner equations 
\beq \label{inner1}
\wapar_{in}'' = \frac{i g \lambda }{ 2 \epsilon \hx }\wphi_{in}'' ,
\eeq
\beq
g \left(\wapar_{in} -  \frac{d_e^2}{\epsilon^2} \wapar_{in}''\right) + i \frac{ 2 \epsilon \hx}{\lambda}\wphi_{in} - i \frac{\rho_s^2 2 \hx}{\lambda \epsilon}\wphi_{in}''=0. \label{inner2}
\eeq
We introduce the real-valued displacement function 
\beq
\xi_{in} = -\frac{i}{g} \wphi_{in},
\eeq
and injecting (\ref{inner1}) in (\ref{inner2}), we obtain the layer equation
\beq \label{layerequation}
\frac{ \xi_{in}''}{\epsilon^2} -\frac{2 \epsilon \hx }{ \lambda \rs^2 \left( \frac{g^2 d_e^2}{\rs^2} + \frac{4 \epsilon^2 \hx^2 }{\lambda^2} \right) } \left( \frac{ \epsilon \hx}{2 \lambda} \xi_{in} -1 \right) =0,
\eeq
where we used the constant $\psi$ approximation, which, we recall, consists in approximating $\wapar_{in} \sim 1$ close to $x=0$. 
In order to solve (\ref{layerequation}) we  will assume 
\beq \label{assumption}
g d_e \ll \rho_s^2 \ll 1,
\eeq
and will make use of a second asymptotic matching inside the inner region. We will have indeed two boundary layers at $x=0$, defining two spatial regions in which the equations can be solved. A boundary layer exists at the scale $\epsilon_1 = \rho_s$ and a second one at a smaller scale, for $\epsilon_2 = \frac{g d_e}{\rho_s}$.

In the first layer we use
\beq \label{rescflayer}
\epsilon=\epsilon_1 = \rho_s, \quad \xi_{in}= \frac{\hxi}{\epsilon_1},
\eeq
where $\hxi$ is the rescaled displacement function. This choice for $\epsilon$ yields a distinguished limit allowing to retain the maximum number of terms in Eq. (\ref{layerequation}), as $\epsilon \rightarrow 0$, accounting for the condition (\ref{assumption}), which allows to neglect the term $g^2 d_e^2/\rho_s^2$ in the denominator of Eq. (\ref{layerequation}). 
We restrict our study to the case of negligible FLR effects in the inner region, which implies that $\rho_e \ll \epsilon_1$. This condition ensures that the terms responsible for the electron FLR effects remain smaller than those responsible for the effects of electron inertia. \\
The rescaling leads to the layer equation 
\beq \label{eqin1}
\hxi'' - \hxi = - \frac{2 \lambda }{\hx}.
\eeq
The solution of Eq. (\ref{eqin1}) is 
\beq
\hxi = \frac{\lambda }{4} e^{ \hx} E_1( \hx) + \frac{\lambda }{4} e^{-  \hx} \left( Ei( \hx) - \frac{ \lambda  g d_e}{\rho_s^2}\frac{\pi}{2} \right). \label{solutionflayer}
\eeq
Where we already fixed the constants of integration in order to ensure $\lim_{z \rightarrow + \infty} \txi =0$ and to ensure the matching with the solution in the second layer.
In (\ref{solutionflayer}) we used the expression of the exponential integral functions 
\begin{equation}
     E_1(x) = \int_x^{+\infty}\frac{e^{-t}}{t}dt, \quad \quad \mbox{and} \quad \quad  Ei(x) = \int_{-\infty}^x \frac{e^{t}}{t}dt. \quad \quad \mbox{for} \quad  x>0,
\end{equation}

\subsection{Inner region : second boundary layer}

In the second layer, where $\hat{x} \sim g d_e /\rho_s^2$, the solution (\ref{solutionflayer}) is no longer valid. Therefore, in the second layer, we perform the following rescaling
\beq 
\epsilon=\epsilon_2 = \frac{g d_e}{\rho_s}, \quad \xi_{in}= \frac{d_e}{\rho_s^3}\bxi,
\eeq
and introduce the second inner variable $\bar{x}=x/\epsilon_2$ (so that $\hat{x}=(g d_e/ \rho_s^2) \bar{x}$).
Since we are at an even smaller spatial scale than that of the previous layer, we emphasize the condition of neglecting the FLR effects also in this second inner layer, i.e. $\rho_e \ll \epsilon_2 $.\\
Considering our assumption (\ref{assumption}), the equation (\ref{layerequation}) becomes,
\beq\label{eqin2}
\bxi'' + \frac{2 \barx }{\lambda\left( 1 + \frac{4 \barx^2}{\lambda^2}\right)} = 0. 
\eeq
The solution of Eq. (\ref{eqin2}), written bellow, in terms of the variables $\hx$ and $ \hxi$ reads
\beq
\begin{split} \label{solex}
\hxi(\hx) = \lambda    \Bigg( 1-\gamma_E+ \frac{\lambda  g d_e }{2 \rho_s^2}\frac{\pi}{2}+ & \log \Bigg( \frac{\rho_s^2}{  g d_e}\Bigg) \Bigg) \hx - \lambda^2 \frac{g d_e}{4 \rho_s^2}\arctan\Bigg(\frac{\rho_s^2 2 \hx}{  g d_e  \lambda} \Bigg)\\
&- \frac{\lambda }{4}\log\Bigg( \Big( \frac{\rho_s^2\hx}{g d_e}\Big)^2+  \frac{\lambda }{q}\Bigg) \hx.
\end{split}
\eeq
This solution satisfies the  boundary condition $\hat{\xi} (0) =0$, descending from the requirement of $\tilde{\phi}$ being an odd function.
In Eq. (\ref{solex}) $\gamma_E$ is the Euler constant.

\subsection{$\Delta'$ matching}
 
We add the following matching condition concerning the derivatives of the solutions:
\begin{equation} \label{intA''}
    \Delta'= \frac{1}{\epsilon_1} \int^{\infty}_{-\infty} \wapar''_{in} d \hx.
\end{equation}
Using the relations (\ref{inner1}) and (\ref{eqin2}) and using the variables $\hx$ and $\hxi$ we write
\beq \label{int}
\Delta'= \frac{2 g^2}{ \rho_s^3} \int^{+ \infty}_0  \frac{\left( 1 - \frac{2 \hx}{\lambda}\hxi\right)}{ \left( \frac{g^2 d_e^2}{ \rho_s^4} + \frac{4 \hx^2}{  \lambda^2}\right) } d\hx .
\eeq
We separate the integral referring to the second term on the right-hand side of Eq. (\ref{int})    in two parts, one from $0$ to $\sigma$ and one from $\sigma$ to $+\infty$, with $\sigma$ a parameter constrained in the overlap region such that 
\beq \label{overlapregion} 
\frac{g d_e}{ \rho_s^2} \ll \sigma \ll \frac{1}{\log\left( \frac{g d_e}{ \rho_s^2} \right) } .
\eeq
We also recall that $\frac{g d_e}{ \rho_s^2} \ll 1$ is our assumption (\ref{assumption}). Equation (\ref{int}) can then be rewritten as
\begin{equation}
\begin{split}\label{troisintegrales}
    & \Delta'= \frac{2 g^2}{ \rho_s^3} \int^{+ \infty}_0  \frac{1}{ \left( \frac{g^2 d_e^2}{ \rho_s^4} + \frac{4 \hx^2}{  \lambda^2}\right) } d\hx - \frac{4 g^2}{\lambda \rho_s^3 } \int^{\sigma}_0 \frac{\hx \hxi }{ \left( \frac{g^2 d_e^2}{ \rho_s^4} + \frac{4 \hx^2}{  \lambda^2}\right) } d\hx  - \frac{4 g^2}{\lambda \rho_s^3 } \int^{\infty}_\sigma \frac{\hx \hxi }{ \left( \frac{g^2 d_e^2}{ \rho_s^4} + \frac{4 \hx^2}{  \lambda^2}\right) } d\hx.\\
    & =\frac{g \lambda \pi  }{2 d_e  \rho_s} + W_2 + W_1.
\end{split}
\end{equation}
We calculate the expression (\ref{troisintegrales}) accurate to $g^2/ \rho_s^3$ so smaller terms are neglected (the next higher term is of order  $ \frac{g^2}{ \rho_s^3} \sigma \log \frac{g d_e}{ \rho_s^2} $ and thanks to the constraint (\ref{overlapregion}) we have $\sigma \log \frac{g d_e}{ \rho_s^2}\ll1$).

 In the interval between $\sigma$ and $+ \infty$, we use the hypothesis (\ref{assumption}), given by $g d_e \ll \rho_s^2 \ll 1$ to simplify the denominator. 
\begin{equation}
\begin{split}
    W_1 & = - \frac{4 g^2}{\lambda \rho_s^3 } \int^{\infty}_\sigma \frac{\hx \hxi }{ \left( \frac{g^2 d_e^2}{ \rho_s^4} + \frac{4 \hx^2}{  \lambda^2}\right) } d\hx.\\
    & = - \frac{g^2}{\rho_s^3 } \int^{\infty}_\sigma \frac{\hx }{ \left( \frac{g^2 d_e^2}{ \rho_s^4} + \frac{4 \hx^2}{  \lambda^2}\right) } \left(  e^{ \hx} E_1( \hx) + e^{-  \hx} \left( Ei( \hx) - \frac{ \lambda  g d_e}{\rho_s^2}\frac{\pi}{2} \right)\right) d\hx.\\
    & =  - \frac{\lambda^2 g^2 }{4\rs^3} \int^{\infty}_\sigma \frac{1}{ \hx} \left(  e^{ \hx} E_1( \hx) + e^{-  \hx} Ei( \hx) \right) d\hx + \frac{\lambda^3 g^3 d_e }{ 4\rs^5} \frac{\pi}{2} \int^{\infty}_\sigma \frac{e^{-  \hx}}{ \hx}  d\hx.
\end{split}
\end{equation}
Using the identity  $e^{u} E_1(u)  + e^{-u} Ei(u)  =  2 \int_0^{\infty} \frac{u}{u^2 + t^2} \sin(t) dt $ (from \cite{Gel69} (id. 22 Tab. 3.3)) and knowing that $\Gamma (0, \sigma)= \int^{\infty}_\sigma \frac{e^{-  \hx}}{ \hx} d \hat{x}$  is the incomplete gamma function whose dominant contribution, as $\sigma \rightarrow 0^+$, is $\log(\sigma)$, we obtain

\begin{equation}
\begin{split}
    W_1 & =  \frac{\lambda^2 g^2 }{ \rs^3} \left( \int_{0}^{\infty} \int_{ \sigma}^{\infty} \frac{\sin(t)}{\hx^2 + t^2} \, d\hx  \, dt  + O \left(\frac{g d_e}{\rho_s^2}\log(\sigma)\right) \right) d\hx,\\
\end{split}
\end{equation}
 when $ \sigma \rightarrow 0^{+}$ and $g d_e/(\rho_s^2\sigma) \rightarrow 0^{+}$. Focusing now on the remaining double integral,
\begin{equation}
\begin{split}
 \int_{0}^{\infty} \int_{\sigma}^{\infty} \frac{\sin(t)}{\hx^2 + t^2} \, d\hx  \, dt   & =\int_{0}^{\infty} \sin(t) \frac{\arctan(\hx/t)}{t} \Big{\rvert}_{ \sigma}^{\infty}  \, dt \\
   & =\frac{\pi}{2} \int_{0}^{\infty} \frac{\sin(t)}{t} \, dt - \int_{0}^{\infty} \frac{\sin(t)}{t} \arctan(\sigma /t) \, dt. 
  \end{split}
\end{equation}
We can prove that the second term is negligible when $ \sigma \rightarrow 0^{+}$ by introducing a new small parameter $\kappa$ such as $ \sigma \ll \kappa \ll 1$, splitting the integral into the sum of an integral from $0$ to $\kappa$ with an integral from $\kappa$ to $+\infty$, and using that in the region $0<t<\kappa$, $\arctan( \sigma /t) < \frac{\pi}{2} $ and $\sin(t) \sim t$ and in the region $\kappa < t $, one has $\arctan(\sigma /t) \sim ( \sigma /t )$. We thus obtain 
\begin{equation}
\begin{split}
    W_1 &= - \frac{\lambda^2 g^2 }{ 4\rs^3} \left( \frac{ \pi^2}{2} +  O \left(\frac{g d_e}{\rho_s^2}\log(\sigma)\right)\right),
   \label{}
\end{split}
\end{equation}
 when $ \sigma \rightarrow 0^{+}$ and $g d_e/(\rho_s^2\sigma) \rightarrow 0^{+}$.\\
It is then possible to show, using (\ref{overlapregion}) and (\ref{assumption}) that 
\begin{equation}
\begin{split}
  W_2 & =  O\left( \frac{g d_e}{\rho_s^2}\log \left(\frac{g d_e}{\rho_s^2}\right)\right) + O\left(\frac{g d_e}{\rho_s^2}\log\left(\sigma\right)\right)  + O \left( \sigma \log \left(\sigma\right)\right)  + O\left( \sigma \log \left(\frac{g d_e}{\rho_s^2}\right)\right) ,
\end{split} 
\end{equation}
 when $ \sigma \rightarrow 0^{+}$ and $g d_e/(\rho_s^2\sigma) \rightarrow 0^{+}$.\\
 Summing all the leading order terms and neglecting the higher order contributions, we obtain the dispersion relation
 \beq \label{disp}
 \Delta' = \frac{g \lambda \pi  }{2 d_e  \rho_s} - \frac{g^2 \lambda^2}{ 4 \rho_s^{3}}\frac{\pi^2}{2}.
\eeq
It is possible, in virtue of (\ref{assumption}), to verify that the second term on the right-hand side of Eq. (\ref{disp}) is smaller than the first one ($g/(d_e  \rho_s) \gg g^2 /  \rho_s^3$). \\
Retaining only the first term in Eq. (\ref{disp}) gives the growth rate predicted by \cite{Por91} and corresponding to the dispersion relation (\ref{smalldelta'}).
When taking into account the corrective term, we obtain the expression for the growth rate
\beq 
\gamma_u = 2 k_y \left( \frac{ \rho_s ^2}{\pi  d_e  \lambda }-\frac{\rho_s^{3/2}\sqrt{\rho_s - 2 d_e^2 \Delta'}}{\pi 
   d_e  \lambda } \right), \label{corrected2}
\eeq
corresponding to Eq. (\ref{corrected}). We remark that, because of the parity properties we required on $\tilde{\phi}$ and $\tilde{A}$, the growth rate $\gamma_u$ has to be real, which enforces a further condition of validity, corresponding to

\begin{equation}
\rho_s \geq 2 d_e^2 \Delta '.
\end{equation}
We performed high precision tests to verify the corrective term of the dispersion relation (\ref{corrected2}). 

%



\bibliographystyle{jpp}

\bibliography{Granier}

\end{document}